# Thermo-micro-mechanical simulation of bulk metal forming processes

S. Amir H. Motaman*,[a], Konstantin Schacht[a], Christian Haase[a], Ulrich Prahl[a,b]

[a]Steel Institute, RWTH Aachen University, Intzestr. 1, D-52072 Aachen, Germany
[b]Institute of Metal Forming, TU Bergakademie Freiberg, Bernhard-von-Cotta-Str. 4, D-09599 Freiberg, Germany

ABSTRACT

The newly proposed microstructural constitutive model for polycrystal viscoplasticity in cold and warm regimes (Motaman and Prahl, 2019), is implemented as a microstructural solver via user-defined material subroutine in a finite element (FE) software. Addition of the microstructural solver to the default thermal and mechanical solvers of a standard FE package enabled coupled thermo-micro-mechanical or thermal-microstructural-mechanical (TMM) simulation of cold and warm bulk metal forming processes. The microstructural solver, which incrementally calculates the evolution of microstructural state variables (MSVs) and their correlation to the thermal and mechanical variables, is implemented based on the constitutive theory of isotropic hypoelasto-viscoplastic (HEVP) finite (large) strain/deformation. The numerical integration and algorithmic procedure of the FE implementation are explained in detail. Then, the viability of this approach is shown for (TMM-) FE simulation of an industrial multistep warm forging.

## Contents

* Corresponding author. Tel: +49 241 80 90133; Fax: +49 241 80 92253
  Email: seyedamirhossein.motaman@iehk.rwth-aachen.de

**Nomenclature**

*Symbol*    *Description*

| Symbol | Description | Units |
|---|---|---|
| $b$ | Burgers length (magnitude of Burgers vector) | [m] |
| **B** | Left Cauchy-Green deformation tensor | [-] |
| $\mathcal{B}$ | Continuum body | [-] |
| $c$ | Constitutive parameter associated with probability amplitude of dislocation processes | [-] |
| **C** | Right Cauchy-Green deformation tensor | [-] |
| $\mathbb{C}$ | Fourth-order stiffness operator/tensor | [Pa] |
| **D** | Rate of deformation tensor | [s$^{-1}$] |
| $E$ | Elastic/Young's modulus | [Pa] |
| **f** | Volumetric body force vector | [N.m$^{-3}$] |
| **F** | Deformation gradient tensor | [-] |
| $G$ | Shear modulus | [Pa] |
| $H$ | Tangent modulus | [Pa] |
| **I** | Unit/identity (second-order) tensor | [-] |
| $\mathbb{I}$ | Fourth-order unit tensor | [-] |
| $J$ | Jacobian of the deformation map | [-] |
| $K$ | Bulk modulus | [Pa] |
| **L** | Velocity gradient tensor | [s$^{-1}$] |
| $m$ | Strain rate sensitivity parameter | [-] |
| $M$ | Taylor factor | [-] |
| **n** | Surface outward normal (unit) vector | [-] |
| **N** | Yield surface normal tensor | [-] |
| **O** | Zero (second-order) tensor | [-] |
| $q$ | Volumetric heat generation | [J.m$^{-3}$] |
| $r$ | Temperature sensitivity coefficient | [-] |
| $R$ | Residual function in Newton-Raphson scheme | |
| **R** | Polar (rigid-body) rotation tensor | [-] |
| $s$ | Temperature sensitivity exponent | [-] |
| $s$ | Stochastic/nonlocal microstructural state set (a set containing all the MSVs) | |
| $t$ | Time | [s] |
| **t** | Traction vector | [Pa] |
| $T$ | Temperature | [K] |
| **u** | Displacement vector | [m] |
| **U** | Right stretch tensor | [-] |
| $\boldsymbol{v}$ | Velocity vector | [m.s$^{-1}$] |
| **V** | Left stretch tensor | [-] |
| $w$ | Volumetric work | [J.m$^{-3}$] |
| **W** | Spin tensor | [s$^{-1}$] |
| $\boldsymbol{x}$ | Position vector (spatial coordinate) | [m] |
| $\alpha$ | Dislocation interaction strength/coefficient | [-] |
| $\beta$ | Dissipation factor, efficiency of plastic dissipation, or Taylor–Quinney coefficient | [-] |
| $\varepsilon$ | Mean/nonlocal (normal) strain | [-] |
| **ε** | Logarithmic/true strain tensor | [-] |
| $\theta$ | Plastic/strain hardening | [Pa] |
| $\varphi$ | Viscous/strain-rate hardening | [Pa.s] |
| $\phi$ | Yield function | |
| $\chi$ | Tolerance | [-] |
| $\psi$ | Flow potential | [Pa] |
| $\kappa$ | Material constant associated with dissipation factor | [-] |
| $\dot{\lambda}$ | Consistency parameter or plastic multiplier | [s$^{-1}$] |
| **Λ** | Rotation tensor | [-] |
| $\nu$ | Poisson's ratio | [-] |
| $\rho$ | Dislocation density | [m$^{-2}$] |
| $\varrho$ | Mass density | [kg.m$^{-3}$] |
| $\sigma$ | Mean/nonlocal (normal) stress | [Pa] |
| **σ** | Cauchy stress tensor | [Pa] |
| **ω** | Spatial skew-symmetric tensor associated with the rotation tensor | [s$^{-1}$] |

| *Index* | *Description* |
|---|---|
| ac | Accumulation |
| an | Annihilation |
| corr | Corrected |
| *d* | Deviatoric/isochoric |
| eff | Effective |
| gn | Generation |
| *h* | Hydrostatic |
| *i* | Immobile |
| *{k}* | Newton-Raphson iteration index, previous Newton-Raphson iteration step |
| *{k+1}* | Current Newton-Raphson iteration step |
| *m* | Mobile, melt |
| min | Minimum |
| *(n)* | Time increment index, previous time increment, beginning of the current time increment |
| *(n+1)* | Current time increment/step, end of the current time increment |
| nc | Nucleation |
| *c* | Cell |
| *p* | Plastic |
| rm | Remobilization |
| tr | Trapping (locking and pinning) |
| trial | Trial step |
| *v* | Viscous (subscript), volumetric (superscript) |
| *w* | Wall |
| *x* | Cell, wall, or total ($x = c, w, t$) |
| *y* | Mobile, immobile, or total ($y = m, i, t$), yield/flow |
| *z* | Dislocation process ($z = $ gn, an, ac, tr, nc, rm) |
| $0$ | Reference, initial/undeformed state |
| $\nabla$ | Objective/material rate of a tensor |
| $\hat{\ }$ | Normalized/dimensionless ($\hat{x} = \frac{x}{x_0}$) |
| $\smile$ | Function |
| $\sim$ | Statistical mean/average |
| $\overline{\ }$ | Equivalent |
| $=$ | Boundary |
| $\underline{\ }$ | Corotational representation of a tensor (rotated to the corotational basis) |

## 1. Introduction

Metal forming processes can be considered as large hypoelasto-viscoplastic deformation under complex varying thermo-mechanical boundary conditions. Moreover, viscoplastic flow of polycrystalline metallic materials is one of the long-standing challenges in classical physics due to its tremendous complexity; and for its accurate continuum description, complex microstructural constitutive modeling is essential.

Microstructural/physics-based material modeling offers the opportunity to enhance the understanding of complex industrial metal forming processes and thus provides the basis for their improvement and optimization. In our previous work (Motaman and Prahl, 2019), the significance of microstructural constitutive models for polycrystal viscoplasticity was pointed out. Application of microstructural state variables (MSVs) including different types of dislocation density was suggested rather than non-measurable virtual internal state variables (ISVs) such as accumulated plastic strain which is not a suitable measure, particularly in complex thermo-mechanical loading condition (varying temperature, strain rate) where history effects are more pronounced (Follansbee and Kocks, 1988; Horstemeyer and Bammann, 2010). However, almost every metal forming simulation performed in industry for design and optimization purposes, apply empirical constitutive models which are based on the accumulated plastic strain as their main ISV. In the last two decades, extensive research in the field of numerical simulation of industrial bulk metal forming has been aimed towards investigation of (thermo-) mechanical aspects of the process such as tools shape and wear, forming force, preform shape, material flow pattern and die filling, etc. (Choi et al., 2012; Guan et al., 2015; Hartley and Pillinger, 2006; Kim et al., 2000; Lee et al., 2013; Ou et al., 2012; Sedighi and Tokmechi, 2008; Vazquez and Altan, 2000; Xianghong et al., 2006; Zhao et al., 2002).

Microstructure of the deforming material and its mechanical properties evolve extensively during metal forming processes. Evolution of microstructure and mechanical properties of the deforming metal directly affects its deformation behavior and consequently the forming process itself as well as in-service performance of the final product. Therefore, in addition to thermo-mechanical simulation of forming processes (simulation of evolution of continuous thermo-mechanical field variables), computation of microstructure and properties evolution of the deforming part by means microstructural state variables through a fully coupled thermo-micro-mechanical (TMM) simulation is of paramount importance. Since process, material, microstructure and properties are highly entangled, resorting to cost-effective simultaneous inter-correlated simulation of process, microstructure and properties facilitates and ensures their efficient and robust design. Currently the literature lacks TMM simulation of complex industrial metal forming processes. Nonetheless, a few instances can be found for TMM simulation of laboratory scale metal forming processes using semi-physical models (Álvarez Hostos et al., 2018; Bok et al., 2014).

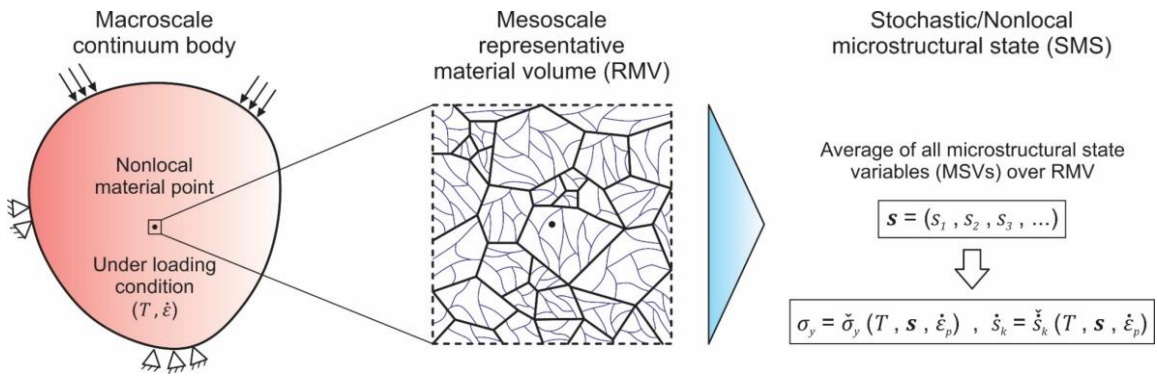

**Fig. 1.** Schematic relation among macroscale continuum body under thermo-mechanical loading, mesoscale representative material volume, and nonlocal microstructural state (Motaman and Prahl, 2019).

The microstructural constitutive models based on continuum microstructure dynamics (CMD) which include continuum dislocation dynamics (CDD) are formulated at macro level, so that the nonlocal MSVs at each macroscale material point in a continuum body are calculated for a (virtual) representative material volume (RMV) around the point based on the evolution/kinetics equations that have physical background, as shown in Fig. 1. The set ***s*** containing all the MSVs is known as the stochastic/nonlocal microstructural state (SMS).

The main objective of the present paper is to show how the microstructural constitutive models based on CDD (as a subset of CMD) can be practically invoked in actual industrial metal forming simulations. The cost of thermo-micro-mechanical (TMM) simulations performed using the applied microstructural constitutive model is in the same range that is offered by common empirical constitutive models. However, since the microstructural models account for the main microstructural processes influencing the material response under viscoplastic deformation, they have a wide range of usability and validity, and can be used in a broad spectrum of deformation parameters (strain rate and temperature). In industrial metal forming processes, polycrystalline materials usually undergo a variety of loading types and parameters; thus, history-dependent microstructural constitutive models are much more suitable and robust for comprehensive simulations of complex industrial metal forming processes. Hence, implementation of the microstructural solver as a user-defined material subroutine in a commercial FE software package and coupling it with the FE software's default mechanical and thermal solvers enables performing realistic TMM simulations of the considered metal forming process chain in order to optimize the process parameters. Interaction among mechanical, thermal and microstructural solvers and their associated fields, together with the initial and boundary conditions and thermo-micro-mechanical properties in fully coupled TMM-FE simulations is shown in Fig. 2.

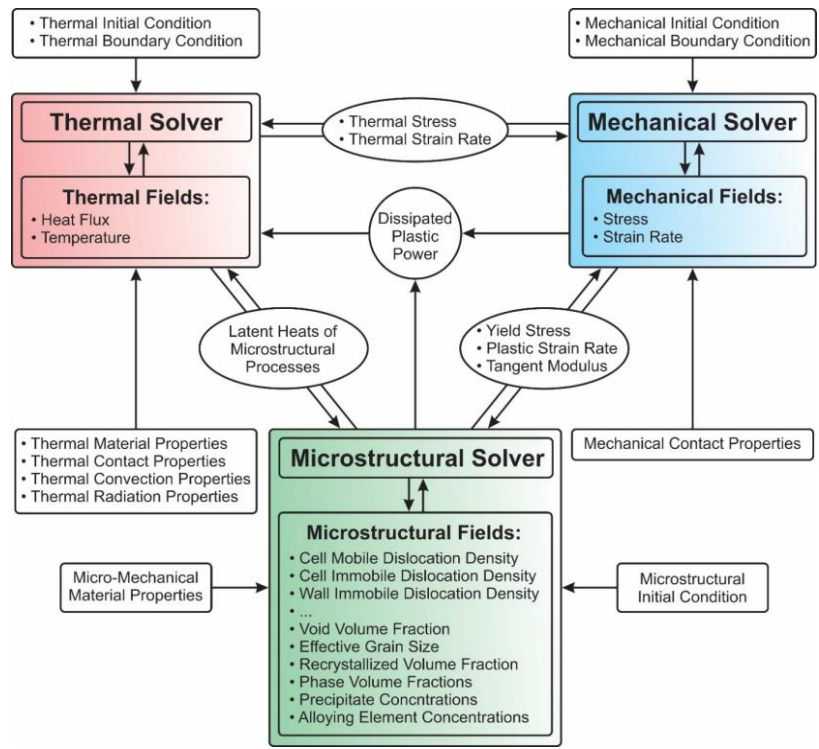

**Fig. 2.** Interaction among mechanical, thermal and microstructural solvers and fields, initial and boundary conditions, and thermo-micro-mechanical properties in fully coupled TMM-FE simulations.

Industrial metal forming processes with respect to temperature are categorized in the following regimes/domains:
- cold regime: cold metal forming processes are conducted in the temperature range starting from room temperature to slightly above it; the maximum temperature in the cold regime is normally characterized by temperatures above which diffusion controlled dislocation mechanisms such as dislocation climb and pinning become dominant (approximately $T < 0.3\ T_m$, where $T$ is the absolute temperature; and $T_m$ is the melting absolute temperature) (Galindo-Nava and Rae, 2016);
- warm regime: warm viscoplastic flow of crystalline materials occurs above the cold but below the hot temperature regime (approximately $0.3\ T_m < T < 0.5\ T_m$) (Berisha et al., 2010; Doherty et al., 1997; Sherby and Burke, 1968); and

- hot regime: hot metal forming processes are carried out above the warm temperature regime. They are characterized by at least one of the hot/extreme microstructural processes such as recrystallization, phase transformation, notable precipitate processes, etc. (roughly $0.5\,T_m < T < T_m$).

Strain rate has different regimes as well, however, independent from the material (Field et al. (2004)):

- creep or static: $\dot{\varepsilon} < 10^{-4}\,\text{s}^{-1}$ (where $\dot{\varepsilon}$ is the strain rate);
- quasi-static: $10^{-4}\,\text{s}^{-1} \leq \dot{\varepsilon} < 10^{-2}\,\text{s}^{-1}$;
- intermediate-rates: $10^{-2}\,\text{s}^{-1} \leq \dot{\varepsilon} \leq 10\,\text{s}^{-1}$;
- dynamic: $10\,\text{s}^{-1} < \dot{\varepsilon} \leq 10^{3}\,\text{s}^{-1}$; and
- shock/highly-dynamic: $\dot{\varepsilon} > 10^{3}\,\text{s}^{-1}$.

The microstructural constitutive model proposed by Motaman and Prahl (2019), has been validated for cold and warm regimes. Moreover, its validity has been verified for the intermediate-rates regime as well, at which most of the industrial bulk metal forming processes are being carried out. In this paper, that constitutive model has been utilized for FE simulation of an actual industrial warm forging process of a bevel gear for automotive applications, made of the ferritic-pearlitic case-hardenable steel 20MnCr5. This particular steel grade which is currently used extensively in industrial bulk metal forming processes in different temperature regimes, has been investigated in hot regime by the recent works of Puchi-Cabrera et al. (2013), (2014) as well as in cold and warm regimes by Brnic et al. (2014) and Motaman and Prahl (2019).

Generally, bulk forming of textureless (randomly oriented grains) undeformed/as-built/annealed polycrystalline metallic materials such as most of the forging and extrusion steel grades can be considered isotropic. Therefore, since deformation of metallic crystalline materials is categorized under HEVP finite strain/deformation, in this paper first the continuum finite strain theory of isotropic HEVP is reformulated in the format of rate equations (without using accumulated strain scalars and tensors). The constitutive equations in corotational configuration are then numerically integrated using various schemes. Finally, the described algorithmic procedure, which is implemented as user-defined material subroutines in ABAQUS, is applied to a bulk metal forming process: industrial multistep warm forging of a bevel gear shaft for automotive applications.

## 2. Continuum finite strain: isotropic hypoelasto-viscoplasticity (HEVP)

### 2.1. Basic kinematics

Consider $\mathcal{B}$ define the current configuration of a continuum body at time $t$, and $\mathcal{B}_0$ the reference, initial or undeformed configuration at the initial time $t = t_0$, where $t$ is time. Let $\boldsymbol{x}_0 \in \mathcal{B}_0$ be the initial position of particle $P$ in the reference configuration $\mathcal{B}_0$, and $\boldsymbol{x} \in \mathcal{B}$ the position of $P$ in the current configuration $\mathcal{B}$. The motion and deformation of the body is defined by a smooth time-dependent mapping $\breve{\chi}_t : \mathcal{B}_0 \to \mathcal{B}$, so that $\boldsymbol{x} = \breve{\chi}(\boldsymbol{x}_0, t)$. Accordingly, the deformation gradient tensor ($\mathbf{F}$) is defined as:

$$\mathbf{F} \equiv \nabla_0 \boldsymbol{x} = \frac{\partial \boldsymbol{x}}{\partial \boldsymbol{x}_0}\,; \tag{1}$$

where $\nabla_0 \equiv \frac{\partial}{\partial \boldsymbol{x}_0}$ is the material gradient operator. This tensor transforms an infinitesimal material vector $\mathrm{d}\boldsymbol{x}_0 \in \mathcal{B}_0$ in the initial configuration $\mathcal{B}_0$, into the corresponding spatial vector $\mathrm{d}\boldsymbol{x} \in \mathcal{B}$ in the current configuration $\mathcal{B}$:

$$\mathrm{d}\boldsymbol{x} = \mathbf{F}\,\mathrm{d}\boldsymbol{x}_0\,. \tag{2}$$

Further, the displacement and velocity vector fields are defined as follows:

$$\boldsymbol{u} \equiv \boldsymbol{x} - \boldsymbol{x}_0 = \breve{\chi}(\boldsymbol{x}_0, t) - \boldsymbol{x}_0 = \boldsymbol{x} - \breve{\chi}^{-1}(\boldsymbol{x}, t)\,; \tag{3}$$

$$\boldsymbol{v} \equiv \frac{\partial \boldsymbol{u}}{\partial t} = \frac{\partial \boldsymbol{x}}{\partial t}\,. \tag{4}$$

Inserting Eq. (3) to Eq. (1) yields:

$$\mathbf{F} = \mathbf{I} + \frac{\partial \boldsymbol{u}}{\partial \boldsymbol{x}_0} = \mathbf{I} + \nabla_0 \boldsymbol{u}\,. \tag{5}$$

Furthermore, the velocity gradient tensor (**L**) is the spatial derivative of $\boldsymbol{v}$ which is given by:

$$\mathbf{L} \equiv \nabla \boldsymbol{v} = \frac{\partial \boldsymbol{v}}{\partial \boldsymbol{x}} = \dot{\mathbf{F}} \mathbf{F}^{-1} \ ; \tag{6}$$

where $\nabla \equiv \frac{\partial}{\partial \boldsymbol{x}}$ is the spatial gradient operator. The velocity gradient is decomposed to its symmetric and skew-symmetric parts, that are respectively known as rate of deformation tensor (**D**) and spin tensor (**W**):

$$\mathbf{L} = \mathbf{D} + \mathbf{W} \ ; \quad \mathbf{D} \equiv \widetilde{\mathrm{sym}}(\mathbf{L}) = \frac{1}{2}(\mathbf{L} + \mathbf{L}^{\mathrm{T}}) \ ; \quad \mathbf{W} \equiv \widetilde{\mathrm{skw}}(\mathbf{L}) = \frac{1}{2}(\mathbf{L} - \mathbf{L}^{\mathrm{T}}) \ . \tag{7}$$

*2.2. Conservation laws*

In the Lagrangian framework (material description) of hypoelasto-viscoplasticity, mass conservation (continuity equation) reads:

$$\varrho_0 = J\varrho \ ; \quad J \equiv \widetilde{\mathrm{det}}(\mathbf{F}) \ ; \tag{8}$$

where $\varrho_0$ and $\varrho$ are mass densities at initial and current configurations, respectively; and $J$ is Jacobian of the deformation map. Here, the angular momentum conservation is satisfied by symmetry of Cauchy stress tensor ($\boldsymbol{\sigma}$):

$$\boldsymbol{\sigma} = \boldsymbol{\sigma}^{\mathrm{T}} \ . \tag{9}$$

The balance of linear momentum for each material point in body $\mathcal{B}$ is established by the following equation:

$$\widetilde{\mathrm{div}}(\boldsymbol{\sigma}) + \boldsymbol{f} = \varrho \dot{\boldsymbol{v}} \ ; \quad \widetilde{\mathrm{div}}(\boldsymbol{\sigma}) \equiv \nabla \cdot \boldsymbol{\sigma} \ ; \tag{10}$$

where $\boldsymbol{f}$ is the volumetric body force vector such as gravitational and magnetic forces. Thus, the body force usually is neglected in metal forming (HEVP) problems ($\boldsymbol{f} = \boldsymbol{0}$). For the derivation of the above-mentioned conservation laws, and also for the finite element discretization of Eq. (10), readers are referred to reference textbooks Belytschko et al. (2014) and/or Zienkiewicz (2005).

*2.3. Initial and boundary conditions*

To complete the problem description, initial and boundary conditions must be provided. The velocity and stress initial conditions for each material particle $P$ at the initial position $\boldsymbol{x}_0 \in \mathcal{B}_0$ and initial time $t_0$ are given by:

$$\breve{\boldsymbol{v}}(\boldsymbol{x}_0, t_0) = \breve{\boldsymbol{v}}_0(\boldsymbol{x}_0) \ ; \quad \boldsymbol{x}_0 \in \mathcal{B}_0 \ ; \tag{11}$$

$$\breve{\boldsymbol{\sigma}}(\boldsymbol{x}_0, t_0) = \breve{\boldsymbol{\sigma}}_0(\boldsymbol{x}_0) \ ; \quad \boldsymbol{x}_0 \in \mathcal{B}_0 \ ; \tag{12}$$

where $\boldsymbol{v}_0$ and $\boldsymbol{\sigma}_0$ are the known initial velocity and stress fields, respectively. Let the boundary of body $\mathcal{B}$ be denoted by $\bar{\bar{\mathcal{B}}}$ and partitioned into disjoint complementary subset boundaries, $\bar{\bar{\mathcal{B}}}_v$ and $\bar{\bar{\mathcal{B}}}_t$ where the essential/velocity/displacement/Dirichlet and natural/traction/Neumann boundary conditions are applied, respectively:

$$\breve{\boldsymbol{v}}(\bar{\boldsymbol{x}}, t) = \bar{\breve{\boldsymbol{v}}}(\bar{\boldsymbol{x}}, t) \ ; \quad \bar{\boldsymbol{x}} \in \bar{\bar{\mathcal{B}}}_v \ ; \tag{13}$$

$$\breve{\boldsymbol{\sigma}}(\bar{\boldsymbol{x}}, t) \cdot \breve{\boldsymbol{n}}(\bar{\boldsymbol{x}}, t) = \bar{\breve{\boldsymbol{t}}}(\bar{\boldsymbol{x}}, t) \ ; \quad \bar{\boldsymbol{x}} \in \bar{\bar{\mathcal{B}}}_t \ ; \tag{14}$$

$$\bar{\bar{\mathcal{B}}}_v \cap \bar{\bar{\mathcal{B}}}_t = \emptyset \ ; \quad \bar{\bar{\mathcal{B}}}_v \cup \bar{\bar{\mathcal{B}}}_t = \bar{\bar{\mathcal{B}}} \ ; \tag{15}$$

where $\bar{\bar{\boldsymbol{v}}}$ and $\bar{\bar{\boldsymbol{t}}}$ are respectively known functions of prescribed boundary velocity and traction vectors; and $\boldsymbol{n}$ is the surface outward normal (unit) vector.

*2.4. Polar decomposition*

The polar decomposition theorem states that any non-singular, second-order tensor can be decomposed uniquely into the product of an orthogonal (rotation) tensor, and a symmetric (stretch) tensor. Since the deformation gradient tensor is a non-singular second-order tensor, the application of polar decomposition theorem to $\mathbf{F}$ implies:

$$\mathbf{F} = \mathbf{R}\mathbf{U} = \mathbf{V}\mathbf{R} \ ; \tag{16}$$

$$\mathbf{R}^{-1} = \mathbf{R}^{\mathrm{T}} \ ; \quad \mathbf{U} = \mathbf{U}^{\mathrm{T}} \ ; \quad \mathbf{V} = \mathbf{V}^{\mathrm{T}} \ ; \tag{17}$$

where $\mathbf{R}$ is the orthogonal polar (rigid-body) rotation tensor; and $\mathbf{U}$ and $\mathbf{V}$ are the right and left stretch tensors, respectively. Hence,

$$\mathbf{U}^2 = \mathbf{C} \equiv \mathbf{F}^{\mathrm{T}}\mathbf{F} \ ; \quad \mathbf{V}^2 = \mathbf{B} \equiv \mathbf{F}\mathbf{F}^{\mathrm{T}} \ ; \tag{18}$$

where $\mathbf{B}$ and $\mathbf{C}$ are left and right Cauchy-Green deformation tensors, respectively.

*2.5. Elasto-plastic decomposition*

The multiplicative elasto-plastic decomposition/split of deformation gradient tensor reads (Kröner, 1959; Lee, 1969; Lee and Liu, 1967; Reina et al., 2018):

$$\mathbf{F} = \mathbf{F}_e\mathbf{F}_p \ ; \quad \widetilde{\det}(\mathbf{F}_p) = 1 \ ; \quad \widetilde{\det}(\mathbf{F}_e) > 0 \ ; \tag{19}$$

where $\mathbf{F}_e$ and $\mathbf{F}_p$ are elastic and plastic deformation gradients, respectively. Combining Eq. (6) and (19) leads to:

$$\mathbf{L} = \mathbf{L}_e + \mathbf{F}_e\mathbf{L}_p\mathbf{F}_e^{-1} \ ; \quad \mathbf{L}_e \equiv \dot{\mathbf{F}}_e\mathbf{F}_e^{-1} \ ; \quad \mathbf{L}_p \equiv \dot{\mathbf{F}}_p\mathbf{F}_p^{-1} \ ; \tag{20}$$

where $\mathbf{L}_e$ and $\mathbf{L}_p$ are elastic and plastic velocity gradients, respectively, that can be additively decomposed to their symmetric and skew-symmetric parts:

$$\mathbf{L}_e = \mathbf{D}_e + \mathbf{W}_e \ ; \quad \mathbf{D}_e \equiv \widetilde{\mathrm{sym}}(\mathbf{L}_e) \ ; \quad \mathbf{W}_e \equiv \widetilde{\mathrm{skw}}(\mathbf{L}_e) \ ; \tag{21}$$

$$\mathbf{L}_p = \mathbf{D}_p + \mathbf{W}_p \ ; \quad \mathbf{D}_p \equiv \widetilde{\mathrm{sym}}(\mathbf{L}_p) \ ; \quad \mathbf{W}_p \equiv \widetilde{\mathrm{skw}}(\mathbf{L}_p) \ ; \tag{22}$$

where $\mathbf{D}_e$ and $\mathbf{D}_p$ are the rates of elastic and plastic deformation gradient tensors, respectively; and $\mathbf{W}_e$ and $\mathbf{W}_p$ are elastic and plastic spin tensors, respectively. Commonly, the deformation of metallic materials is considered hypoelasto-viscoplastic. Thus, generally, it can be assumed that elastic strains (rates) are very small compared to unity and plastic strains (and rates). This restriction results in the following approximation (Nemat-Nasser, 1979):

$$\mathbf{F}_e \approx \mathbf{V}_e \approx \mathbf{U}_e \approx \mathbf{I} \ ; \tag{23}$$

where $\mathbf{I}$ is the second-order unit/identity tensor. From this, Eq. (20) turns to (Green and Naghdi, 1965):

$$\mathbf{L} = \mathbf{L}_e + \mathbf{L}_p \ . \tag{24}$$

Therefore, considering Eqs. (7), (20), (21), (22), (23) and (24) (Dunne, 2011; Khan and Huang, 1995):

$$\mathbf{D} = \mathbf{D}_e + \mathbf{D}_p \ ; \quad \mathbf{W} = \mathbf{W}_e + \mathbf{W}_p \ . \tag{25}$$

In this context, the strain rate measure is the power (work) conjugate of Cauchy stress tensor, and thus is the rate of deformation gradient tensor. Consequently,

$$\dot{\boldsymbol{\varepsilon}} = \dot{\boldsymbol{\varepsilon}}_e + \dot{\boldsymbol{\varepsilon}}_p \ ; \quad \dot{\boldsymbol{\varepsilon}} \equiv \mathbf{D} \ ; \quad \dot{\boldsymbol{\varepsilon}}_e \equiv \mathbf{D}_e \ ; \quad \dot{\boldsymbol{\varepsilon}}_p \equiv \mathbf{D}_p \ ; \tag{26}$$

where $\dot{\boldsymbol{\varepsilon}}$, $\dot{\boldsymbol{\varepsilon}}_e$, $\dot{\boldsymbol{\varepsilon}}_p$ are respectively total, elastic and plastic (logarithmic/true) strain rate tensors.

## 2.6. Corotational formulation

Physically motivated material objectivity/frame-indifference principle demands independence of material properties from the respective frame of reference or observer (Truesdell and Noll, 1965). Constitutive equations of HEVP are formulated in a rotation-neutralized configuration with the aid of local coordinate system/basis that rotates with the material. In this framework, the rotation of the neighborhood of a material point is characterized by orthogonal rotation tensor $\boldsymbol{\Lambda}$, which is subjected to the following evolutionary equation and initial condition (de-Souza Neto et al., 2008; Simo and Hughes, 1998):

$$\dot{\boldsymbol{\Lambda}} = \boldsymbol{\omega}\boldsymbol{\Lambda} \,; \quad \boldsymbol{\Lambda}_0 = \mathbf{I} \,; \tag{27}$$

$$\boldsymbol{\Lambda}^{-1} = \boldsymbol{\Lambda}^{\mathrm{T}} \,; \quad \boldsymbol{\omega} = -\boldsymbol{\omega}^{\mathrm{T}} \,; \tag{28}$$

Where $\boldsymbol{\Lambda}_0$ is initial (at time $t = 0$) $\boldsymbol{\Lambda}$; and $\boldsymbol{\omega}$ is a spatial skew-symmetric (second-order) tensor associated with the rotation tensor. Hence,

$$\boldsymbol{\omega} = \dot{\boldsymbol{\Lambda}}\boldsymbol{\Lambda}^{\mathrm{T}} \,. \tag{29}$$

Therefore, the (symmetric) Cauchy stress tensor ($\boldsymbol{\sigma}$) is rotated to the rotation-neutralized configuration by multiplying it from the left and right with $\boldsymbol{\Lambda}^{\mathrm{T}}$ and $\boldsymbol{\Lambda}$, respectively:

$$\underline{\boldsymbol{\sigma}} = \boldsymbol{\Lambda}^{\mathrm{T}}\boldsymbol{\sigma}\boldsymbol{\Lambda} \quad \Rightarrow \quad \boldsymbol{\sigma} = \boldsymbol{\Lambda}\underline{\boldsymbol{\sigma}}\boldsymbol{\Lambda}^{\mathrm{T}} \,; \tag{30}$$

where $\underline{\phantom{x}}$ denotes corotational representation of a tensor (rotated to the corotational basis). Moreover, the (symmetric) rate of deformation tensor in the corotational configuration reads:

$$\underline{\mathbf{D}} = \boldsymbol{\Lambda}^{\mathrm{T}}\mathbf{D}\boldsymbol{\Lambda} \quad \Rightarrow \quad \mathbf{D} = \boldsymbol{\Lambda}\underline{\mathbf{D}}\boldsymbol{\Lambda}^{\mathrm{T}} \,. \tag{31}$$

Considering Eqs. (27) and (28), time differentiation of the rotated Cauchy stress tensor (Eq. (30)) leads to:

$$\underline{\dot{\boldsymbol{\sigma}}} = \boldsymbol{\Lambda}^{\mathrm{T}}\boldsymbol{\sigma}^{\nabla}\boldsymbol{\Lambda} \quad \Rightarrow \quad \boldsymbol{\sigma}^{\nabla} = \boldsymbol{\Lambda}\underline{\dot{\boldsymbol{\sigma}}}\boldsymbol{\Lambda}^{\mathrm{T}} \,; \tag{32}$$

so that,

$$\boldsymbol{\sigma}^{\nabla} = \dot{\boldsymbol{\sigma}} + \boldsymbol{\sigma}\boldsymbol{\omega} - \boldsymbol{\omega}\boldsymbol{\sigma} \,; \tag{33}$$

where $\boldsymbol{\sigma}^{\nabla}$ is referred to as objective/frame-invariant/material rate of Cauchy stress tensor. In HEVP finite strain, depending on the FE formulation, commonly two members of the family of objective stress rates are considered (Doghri, 2000; Johnson and Bammann, 1984; Mourad et al., 2014):

- Green-Naghdi rate, corresponding to $\boldsymbol{\Lambda} = \mathbf{R}$ (Green and Naghdi, 1971): the rotation is the same as orthogonal polar (rigid-body) rotation tensor $\mathbf{R}$, which can be calculated by tensor Eqs. (16), (17) and (18), using the spectral decomposition (eigen-projection) method.
- Jaumann rate, corresponding to $\boldsymbol{\omega} = \mathbf{W}$: in this case, often the widely used Hughes-Winget approximation (Hughes and Winget, 1980) based on the midpoint rule is applied for calculation of the rotation tensor. The Hughes-Winget formula is valid if the increment of spin tensor ($\Delta t\, \mathbf{W}$) is sufficiently small (adequately small incremental rotations).

For the details of numerical update algorithms of incremental finite rotations associated with the Green-Naghdi and Jaumann rates which are usually based on midpoint method (at the midpoint configuration), readers are encouraged to refer to Simo and Hughes (1998), de-Souza Neto et al. (2008), and/or Belytschko et al. (2014).

## 2.7. Constitutive relation of isotropic HEVP

In HEVP with plastic incompressibility, volume change during deformation is fully elastic and negligible. Therefore, according to the isotropic three-dimensional Hook's law and the material objectivity principle:

$$\boldsymbol{\sigma}^\nabla = \mathbb{C}_e : \dot{\boldsymbol{\varepsilon}}_e = \mathbb{C}_e : \left(\dot{\boldsymbol{\varepsilon}} - \dot{\boldsymbol{\varepsilon}}_p\right);\tag{34}$$

where $\mathbb{C}_e$ is the fourth-order (isotropic) elastic stiffness tensor which is calculated according to:

$$\mathbb{C}_e = 2G\mathbb{I}^d + K\mathbf{I}\otimes\mathbf{I} = 2G\mathbb{I} + \left(K - \frac{2}{3}G\right)\mathbf{I}\otimes\mathbf{I};\tag{35}$$

$$G = \frac{1}{2(1+v)}E;\quad K = \frac{1}{3(1-2v)}E = \frac{2(1+v)}{3(1-2v)}G;\tag{36}$$

$$\mathbb{I}^d \equiv \mathbb{I} - \mathbb{I}^v;\quad \mathbb{I}^v \equiv \frac{1}{3}\mathbf{I}\otimes\mathbf{I};\tag{37}$$

where $E$, $G$, $K$ and $v$ are respectively elastic, shear and bulk moduli and poisson's ratio; and $\mathbb{I}$, $\mathbb{I}^d$ and $\mathbb{I}^v$ being the unit and unit deviatoric and unit volumetric fourth-order tensors, respectively.

*2.8. Deviatoric-volumetric decomposition*

The strain rate tensor can be additively decomposed to deviatoric/isochoric and volumetric parts:

$$\dot{\boldsymbol{\varepsilon}} = \dot{\boldsymbol{\varepsilon}}^d + \dot{\boldsymbol{\varepsilon}}^v;\quad \dot{\boldsymbol{\varepsilon}}^v \equiv \frac{1}{3}\dot{\varepsilon}^v\mathbf{I};\quad \dot{\varepsilon}^v = \mathbf{I}:\dot{\boldsymbol{\varepsilon}}.\tag{38}$$

Given Eq. (37):

$$\dot{\boldsymbol{\varepsilon}}^d = \dot{\boldsymbol{\varepsilon}} - \dot{\boldsymbol{\varepsilon}}^v = \dot{\boldsymbol{\varepsilon}} - \frac{1}{3}\dot{\varepsilon}^v\mathbf{I} = \mathbb{I}^d:\dot{\boldsymbol{\varepsilon}};\quad \dot{\boldsymbol{\varepsilon}}^v = \mathbb{I}^v:\dot{\boldsymbol{\varepsilon}};\tag{39}$$

where superscripts $d$ and $v$ denote deviatoric and volumetric decomposition of the corresponding tensor, respectively. Moreover, the elastic and plastic strain rate tensors can be decomposed to their deviatoric and volumetric parts:

$$\dot{\boldsymbol{\varepsilon}}_e = \dot{\boldsymbol{\varepsilon}}_e^d + \dot{\boldsymbol{\varepsilon}}_e^v;\quad \dot{\boldsymbol{\varepsilon}}_p = \dot{\boldsymbol{\varepsilon}}_p^d + \dot{\boldsymbol{\varepsilon}}_p^v.\tag{40}$$

Hence, considering Eq. (26):

$$\dot{\boldsymbol{\varepsilon}}^d = \dot{\boldsymbol{\varepsilon}}_e^d + \dot{\boldsymbol{\varepsilon}}_p^d;\quad \dot{\boldsymbol{\varepsilon}}^v = \dot{\boldsymbol{\varepsilon}}_e^v + \dot{\boldsymbol{\varepsilon}}_p^v.\tag{41}$$

In pressure-independent plasticity, the volumetric elastic strain rate is responsible for the entire volume change during the elasto-plastic deformation (Aravas, 1987); meaning that:

$$\dot{\boldsymbol{\varepsilon}}_p^v = \mathbf{0}\quad \Rightarrow\quad \dot{\boldsymbol{\varepsilon}}_p = \dot{\boldsymbol{\varepsilon}}_p^d.\tag{42}$$

Thereby,

$$\dot{\boldsymbol{\varepsilon}}^v = \dot{\boldsymbol{\varepsilon}}_e^v = \frac{1}{3}\dot{\varepsilon}_e^v\mathbf{I} = \mathbb{I}^v:\dot{\boldsymbol{\varepsilon}}_e;\quad \dot{\varepsilon}_e^v = \mathbf{I}:\dot{\boldsymbol{\varepsilon}}_e = \dot{\varepsilon}^v = \mathbf{I}:\dot{\boldsymbol{\varepsilon}};\tag{43}$$

$$\dot{\boldsymbol{\varepsilon}}_e^d = \dot{\boldsymbol{\varepsilon}}_e - \dot{\boldsymbol{\varepsilon}}_e^v = \dot{\boldsymbol{\varepsilon}}_e - \frac{1}{3}\dot{\varepsilon}_e^v\mathbf{I} = \mathbb{I}^d:\dot{\boldsymbol{\varepsilon}}_e.\tag{44}$$

Thus, Eqs. (40) and (41) can be rewritten as:

$$\dot{\boldsymbol{\varepsilon}}_e = \dot{\boldsymbol{\varepsilon}}_e^d + \dot{\boldsymbol{\varepsilon}}^v;\quad \dot{\boldsymbol{\varepsilon}}^d = \dot{\boldsymbol{\varepsilon}}_e^d + \dot{\boldsymbol{\varepsilon}}_p.\tag{45}$$

Knowing that the hydrostatic/volumetric parts of objective/material and spatial time derivatives of stress tensor are equal, since they are proportional to the first invariant of stress rate tensors, the objective stress rate tensor is decomposed to its deviatoric and hydrostatic splits as follows:

$$\boldsymbol{\sigma}^{\nabla} = \boldsymbol{\sigma}^{\nabla d} + \dot{\boldsymbol{\sigma}}^{h} ; \tag{46}$$

$$\dot{\boldsymbol{\sigma}}^{h} \equiv \dot{\sigma}^{h}\mathbf{I} = \mathbb{I}^{v}{:}\dot{\boldsymbol{\sigma}} = \mathbb{I}^{v}{:}\boldsymbol{\sigma}^{\nabla} = \boldsymbol{\sigma}^{\nabla h}; \quad \dot{\sigma}^{h} = \frac{1}{3}\mathbf{I}{:}\dot{\boldsymbol{\sigma}} ; \tag{47}$$

$$\boldsymbol{\sigma}^{\nabla d} = \boldsymbol{\sigma}^{\nabla} - \dot{\boldsymbol{\sigma}}^{h} = \boldsymbol{\sigma}^{\nabla} - \dot{\sigma}^{h}\mathbf{I} = \mathbb{I}^{d}{:}\boldsymbol{\sigma}^{\nabla} ; \tag{48}$$

where superscript $h$ denotes hydrostatic contribution of the corresponding tensor. Taking into account Eqs. (34), (35), (37), (43), (44), (46), (47) and (48) results in:

$$\boldsymbol{\sigma}^{\nabla} = \boldsymbol{\sigma}^{\nabla d} + \dot{\boldsymbol{\sigma}}^{h} = 2G\dot{\boldsymbol{\varepsilon}}_{e}^{d} + K\dot{\varepsilon}^{v}\mathbf{I} = 2G\dot{\boldsymbol{\varepsilon}}_{e} + \left(K - \frac{2}{3}G\right)\dot{\varepsilon}^{v}\mathbf{I} ; \tag{49}$$

$$\boldsymbol{\sigma}^{\nabla d} = 2G\dot{\boldsymbol{\varepsilon}}_{e}^{d} ; \quad \dot{\boldsymbol{\sigma}}^{h} = K\dot{\varepsilon}^{v}\mathbf{I} = K\dot{\varepsilon}_{e}^{v}\mathbf{I} . \tag{50}$$

Therefore, by time integration in a fixed arbitrary spatial coordinate system:

$$\boldsymbol{\sigma} = \boldsymbol{\sigma}^{d} + \boldsymbol{\sigma}^{h} ; \tag{51}$$

$$\boldsymbol{\sigma}^{h} = \sigma^{h}\mathbf{I} = \mathbb{I}^{v}{:}\boldsymbol{\sigma} ; \quad \sigma^{h} = \frac{1}{3}\mathbf{I}{:}\boldsymbol{\sigma} ; \tag{52}$$

$$\boldsymbol{\sigma}^{d} = \boldsymbol{\sigma} - \boldsymbol{\sigma}^{h} = \boldsymbol{\sigma} - \sigma^{h}\mathbf{I} = \mathbb{I}^{d}{:}\boldsymbol{\sigma} . \tag{53}$$

## 2.9. Associative isotropic $J_2$ plasticity

The (hydrostatic-) pressure-independent yield criterion for isotropic hardening is adopted through definition of the following yield function ($\phi$):

$$\phi \equiv \bar{\sigma} - \sigma_{y} ; \tag{54}$$

$$\bar{\sigma} \equiv \breve{\sigma}(\boldsymbol{\sigma}) = \sqrt{3\,\breve{J}_{2}(\boldsymbol{\sigma})} = \sqrt{\frac{3}{2}}\|\boldsymbol{\sigma}^{d}\| ; \quad \sigma_{y} \equiv \breve{\sigma}_{y}\bigl(T, \boldsymbol{s}, \dot{\bar{\varepsilon}}_{p}\bigr) . \tag{55}$$

where $\bar{\sigma}$ and $\sigma_y$ are equivalent ($J_2$/von Mises) stress and (scaler) yield stress, respectively; $T$ is the temperature; $\dot{\bar{\varepsilon}}_p$ is the equivalent (von Mises) plastic strain rate; $\|\mathbf{a}\| \equiv \sqrt{\mathbf{a}{:}\mathbf{a}}$ denotes the Euclidean norm of second-order tensor $\mathbf{a}$; and $\breve{J}_2(\boldsymbol{\sigma})$ is the second invariant of deviatoric part of Cauchy stress tensor $\boldsymbol{\sigma}$:

$$\breve{J}_{2}(\boldsymbol{\sigma}) = \frac{1}{2}\|\boldsymbol{\sigma}^{d}\|^{2} \quad \Rightarrow \quad \frac{\partial J_{2}}{\partial \boldsymbol{\sigma}} = \boldsymbol{\sigma}^{d} . \tag{56}$$

In addition, due to invariance of $J_2$, for an arbitrarily rotated Cauchy stress tensor $\underline{\boldsymbol{\sigma}}$:

$$\breve{J}_{2}(\underline{\boldsymbol{\sigma}}) = \breve{J}_{2}(\boldsymbol{\sigma}) \quad \Rightarrow \quad \bar{\sigma} \equiv \breve{\sigma}(\boldsymbol{\sigma}) = \breve{\sigma}(\underline{\boldsymbol{\sigma}}) . \tag{57}$$

In the associative plasticity, the flow potential ($\psi$) is taken as the yield function ($\psi = \phi$), leading to the following associative flow rule (Rice, 1971, 1970):

$$\dot{\boldsymbol{\varepsilon}}_{p} = \dot{\lambda}\frac{\partial \psi}{\partial \boldsymbol{\sigma}} = \dot{\lambda}\frac{\partial \phi}{\partial \boldsymbol{\sigma}} = \dot{\lambda}\mathbf{N} ; \quad \mathbf{N} \equiv \frac{\partial \phi}{\partial \boldsymbol{\sigma}} ; \tag{58}$$

where $\dot{\lambda}$ is the non-negative consistency parameter or viscoplastic multiplier (Simo and Hughes, 1998); and $\mathbf{N}$ is known as flow direction tensor which represents the yield surface normal tensor ($\mathbf{N}$ is not necessarily a unit tenor). Given Eqs. (54), (55), (56) and (58):

$$\mathbf{N} = \frac{3}{2}\frac{\boldsymbol{\sigma}^d}{\bar{\sigma}} \quad \Rightarrow \quad \dot{\boldsymbol{\varepsilon}}_p = \frac{3}{2}\frac{\dot{\lambda}}{\bar{\sigma}}\boldsymbol{\sigma}^d \ ; \tag{59}$$

which sometimes are referred to as the Prandtl-Reuss equations. Taking Euclidean norm from both sides of Eq. (59) leads to:

$$\dot{\lambda} = \sqrt{\frac{2}{3}}\|\dot{\boldsymbol{\varepsilon}}_p\| \ . \tag{60}$$

According to the power (work) equivalence principle, (scaler) volumetric plastic power ($\dot{w}_p$) in multiaxial state can be equally expressed by the equivalent stress and equivalent plastic strain rate ($\dot{\bar{\varepsilon}}_p$):

$$\dot{w}_p = \boldsymbol{\sigma}^d : \dot{\boldsymbol{\varepsilon}}_p = \bar{\sigma}\,\dot{\bar{\varepsilon}}_p \geq 0 \ . \tag{61}$$

Combining Eqs. (59), (60) and (61) results in:

$$\dot{\lambda} = \dot{\bar{\varepsilon}}_p \equiv \check{\bar{\varepsilon}}_p(\dot{\boldsymbol{\varepsilon}}_p) = \sqrt{\frac{2}{3}}\|\dot{\boldsymbol{\varepsilon}}_p\| \ ; \quad \dot{\boldsymbol{\varepsilon}}_p = \dot{\bar{\varepsilon}}_p \mathbf{N} \ . \tag{62}$$

Substituting Eq. (62) into Eq. (59) yields the Levy-Mises flow rule:

$$\dot{\boldsymbol{\varepsilon}}_p = \frac{3}{2}\frac{\dot{\bar{\varepsilon}}_p}{\bar{\sigma}}\boldsymbol{\sigma}^d \ . \tag{63}$$

Finally, formulation is completed by introducing Kuhn-Tucker loading-unloading complementary conditions:

$$\dot{\lambda} \geq 0 \ ; \quad \phi \leq 0 \ ; \quad \dot{\lambda}\phi = 0 \ ; \tag{64}$$

and the consistency condition:

$$\dot{\lambda}\dot{\phi} = 0 \ . \tag{65}$$

Therefore, during viscoplastic deformation ($\dot{\lambda} > 0$), $\dot{\lambda}\phi = 0$ reduces to $\phi = 0$ which is identical to $\bar{\sigma} = \sigma_y$, given Eq. (54). Also, the consistency condition during viscoplastic deformation with isotropic hardening according to yield function defined by Eq. (54), becomes (de-Borst et al., 2014; Wang et al., 1997):

$$\dot{\phi} = \frac{\partial \phi}{\partial \boldsymbol{\sigma}} : \dot{\boldsymbol{\sigma}} - \dot{\sigma}_y = \sqrt{\frac{3}{2}}\mathbf{N} : \dot{\boldsymbol{\sigma}} - \dot{\sigma}_y = 0 \ ; \quad \dot{\sigma}_y = \frac{\partial \sigma_y}{\partial \lambda}\dot{\lambda} + \frac{\partial \sigma_y}{\partial \dot{\lambda}}\ddot{\lambda} \ ; \tag{66}$$

where $\dot{\sigma}_y$ is the viscoplastic hardening rate. In case of associative isotropic $J_2$ plasticity ($\dot{\lambda} = \dot{\bar{\varepsilon}}_p$):

$$\dot{\sigma}_y = \theta\,\dot{\bar{\varepsilon}}_p + \varphi\,\ddot{\bar{\varepsilon}}_p \ ; \quad \theta \equiv \frac{\partial \sigma_y}{\partial \bar{\varepsilon}_p} \ ; \quad \varphi \equiv \frac{\partial \sigma_y}{\partial \dot{\bar{\varepsilon}}_p} \ ; \tag{67}$$

where $\theta$ and $\varphi$ are plastic/strain hardening and viscous/strain-rate hardening, respectively.

*2.10. Corotational representation of constitutive equations*

Taking advantage of corotational formulation (section 2.6), the orthogonality of the rotation tensor ($\boldsymbol{\Lambda}^{-1} = \boldsymbol{\Lambda}^T$), symmetry of Cauchy stress tensor ($\boldsymbol{\sigma} = \boldsymbol{\sigma}^T$), and the isotropy of elastic stiffness tensor ($\underline{\mathbb{C}}_e = \mathbb{C}_e$), the tensor equations described in sections 2.7, 2.8 and 2.9 are form-identical in the corotational configuration but with the spatial tensor variables replaced with their corotational counterparts (Zaera and Fernández-Sáez, 2006). The corotational representation of some of those equations are:

$$\underline{\dot{\boldsymbol{\sigma}}} = \mathbb{C}_e : \underline{\dot{\boldsymbol{\varepsilon}}}_e \; ; \tag{68}$$

$$\underline{\dot{\boldsymbol{\varepsilon}}} = \underline{\dot{\boldsymbol{\varepsilon}}}_e + \underline{\dot{\boldsymbol{\varepsilon}}}_p \; ; \tag{69}$$

$$\underline{\dot{\boldsymbol{\varepsilon}}} = \underline{\dot{\boldsymbol{\varepsilon}}}^d + \frac{1}{3}\dot{\varepsilon}^v \mathbf{I} \; ; \quad \dot{\varepsilon}^v = \mathbf{I} : \underline{\dot{\boldsymbol{\varepsilon}}} = \dot{\varepsilon}^v_e = \mathbf{I} : \underline{\dot{\boldsymbol{\varepsilon}}}_e \; ; \tag{70}$$

$$\underline{\dot{\boldsymbol{\varepsilon}}}_e = \underline{\dot{\boldsymbol{\varepsilon}}}^d_e + \frac{1}{3}\dot{\varepsilon}^v \mathbf{I} \; ; \tag{71}$$

$$\underline{\dot{\boldsymbol{\varepsilon}}}_p = \underline{\dot{\boldsymbol{\varepsilon}}}^d_p = \underline{\dot{\boldsymbol{\varepsilon}}}^d - \underline{\dot{\boldsymbol{\varepsilon}}}^d_e \; ; \tag{72}$$

$$\underline{\dot{\boldsymbol{\sigma}}} = \underline{\dot{\boldsymbol{\sigma}}}^d + \dot{\sigma}^h \mathbf{I} \; ; \quad \underline{\dot{\boldsymbol{\sigma}}}^d = 2G\underline{\dot{\boldsymbol{\varepsilon}}}^d_e \; ; \quad \dot{\sigma}^h = \frac{1}{3}\mathbf{I} : \underline{\dot{\boldsymbol{\sigma}}} = K\dot{\varepsilon}^v \; ; \tag{73}$$

$$\underline{\dot{\boldsymbol{\sigma}}} = 2G\underline{\dot{\boldsymbol{\varepsilon}}}_e + \left(K - \frac{2}{3}G\right)\dot{\varepsilon}^v \mathbf{I} \; ; \tag{74}$$

$$\underline{\boldsymbol{\sigma}} = \underline{\boldsymbol{\sigma}}^d + \sigma^h \mathbf{I} \; ; \quad \sigma^h = \frac{1}{3}\mathbf{I} : \underline{\boldsymbol{\sigma}} \; ; \tag{75}$$

$$\bar{\sigma} \equiv \breve{\bar{\sigma}}(\boldsymbol{\sigma}) = \breve{\bar{\sigma}}(\underline{\boldsymbol{\sigma}}) = \sqrt{\frac{3}{2}} \|\underline{\boldsymbol{\sigma}}^d\| \; ; \tag{76}$$

$$\dot{\bar{\varepsilon}}_p \equiv \breve{\dot{\bar{\varepsilon}}}_p(\dot{\boldsymbol{\varepsilon}}_p) = \breve{\dot{\bar{\varepsilon}}}_p(\underline{\dot{\boldsymbol{\varepsilon}}}_p) = \sqrt{\frac{2}{3}} \|\underline{\dot{\boldsymbol{\varepsilon}}}_p\| \; ; \tag{77}$$

$$\underline{\mathbf{N}} = \frac{3}{2}\frac{\underline{\boldsymbol{\sigma}}^d}{\bar{\sigma}} = \frac{\underline{\dot{\boldsymbol{\varepsilon}}}_p}{\dot{\bar{\varepsilon}}_p} \quad \Rightarrow \quad \underline{\dot{\boldsymbol{\varepsilon}}}_p = \frac{3}{2}\frac{\dot{\bar{\varepsilon}}_p}{\bar{\sigma}}\underline{\boldsymbol{\sigma}}^d \; . \tag{78}$$

The numerical time integration of the above-mentioned corotational representation of constitutive equations and the resultant algorithmic procedure for finite element implementation is explained in section 4.

## 3. Microstructural constitutive model

The microstructural constitutive model for metal isotropic viscoplasticity has the form $\sigma_y = \breve{\sigma}_y(T, \boldsymbol{s}, \dot{\bar{\varepsilon}}_p)$. In the case of cold and warm regimes, the stochastic/nonlocal microstructural state set is $\boldsymbol{s} = \{\rho_{cm}, \rho_{ci}, \rho_{wi}\}$, where $\rho$ is nonlocal dislocation density; subscripts $c$ and $w$ denote cell and wall; and subscripts $m$ and $i$ represent mobile and immobile, respectively. Thus, $\rho_{cm}$, $\rho_{ci}$ and $\rho_{wi}$ are cell mobile, cell immobile and wall immobile dislocation densities, respectively. According to Motaman and Prahl (2019), the microstructural constitutive model for polycrystal viscoplasticity in cold and warm regimes based on continuum dislocation dynamics consists of the following main equations:

$$\sigma_y = \sigma_v + \sigma_p \; ; \quad \sigma_p = \sigma_{pc} + \sigma_{pw} \; ; \tag{79}$$

$$\sigma_{px} = MbG\tilde{\alpha}_x \sqrt{\rho_{xi}} \; ; \quad x = c, w \; ; \tag{80}$$

$$\sigma_v = \sigma_{v0}\dot{\bar{\varepsilon}}_p^{m_v} \; ; \quad \sigma_{v0} \equiv \sigma_{v00}\left[1 + r_v\left(\hat{T} - 1\right)^{s_v}\right] \; ; \quad \hat{T} \equiv \frac{T}{T_0} \; ; \quad \dot{\hat{\bar{\varepsilon}}}_p \equiv \frac{\dot{\bar{\varepsilon}}_p}{\dot{\bar{\varepsilon}}_0} \; ; \quad r_v < 0 \; ; \quad 0 < s_v \leq 1 \; ; \tag{81}$$

where subscripts $v$ and $p$, respectively stand for viscous and plastic; $M$ is the Taylor factor; $b$ is the Burgers length (magnitude of Burgers vector); $\tilde{\alpha}_x$ is the nonlocal interaction strength related to local density and geometrical arrangement of immobile dislocations of cell and wall species ($x = c, w$); $r$ and $s$ are temperature sensitivity coefficient and exponent; $m$ is the strain rate sensitivity parameter; the hat-sign ( $\hat{}$ ) indicates normalization; $T$ is absolute temperature; and here subscript 0 denotes the reference state. Moreover, as a rule, the reference temperature and strain rate are assumed to be the lowest temperature and strain rate in the corresponding investigated regimes, respectively. Combination of shear modulus and mean interaction strengths ($G\tilde{\alpha}_x$ factor in Eq. (80)) and strain rate sensitivity of viscous stress ($m_v$) depend on temperature and strain rate:

$$\widehat{G\tilde{\alpha}}_x = 1 + r_{\alpha x}^G (\hat{T} - 1)^{s_{\alpha x}^G} ; \quad \widehat{G\tilde{\alpha}}_x \equiv \frac{G\tilde{\alpha}_x}{G_0 \tilde{\alpha}_{x0}} ; \quad r_{\alpha x}^G < 0 ; \quad s_{\alpha x}^G > 0 ; \quad x = c, w ; \tag{82}$$

$$\hat{m}_v = \left[1 + r_v^m (\hat{T} - 1)^{s_v^m}\right] \dot{\hat{\varepsilon}}_p^{m_v^m} ; \quad \hat{m}_v \equiv \frac{m_v}{m_{v0}} ; \quad r_v^m, s_v^m \geq 0 ; \tag{83}$$

where $r_{\alpha x}^G$ and $s_{\alpha x}^G$ are temperature sensitivity coefficient and exponent associated with $G\tilde{\alpha}_x$; $m_{v0}$ is the reference (at reference temperature and strain rate) strain rate sensitivity; $r_v^m$ and $s_v^m$ are respectively temperature sensitivity coefficient and exponent associated with strain rate sensitivity of viscous stress; and $m_v^m$ is the strain rate sensitivity parameter associated with strain rate sensitivity of viscous stress. The following equations describe evolution of different types of dislocation densities:

$$\partial_{\bar{\varepsilon}_p} \hat{\rho}_{wi} = \partial_{\bar{\varepsilon}_p} \hat{\rho}_{wi}^{nc} + \partial_{\bar{\varepsilon}_p} \hat{\rho}_{wi}^{ac} - \left(\partial_{\bar{\varepsilon}_p} \hat{\rho}_{wi}^{an} + \partial_{\bar{\varepsilon}_p} \hat{\rho}_{wi}^{rm}\right) ; \quad \hat{\rho}_{xy} \equiv \frac{\rho_{xy}}{\rho_0} ; \quad \begin{cases} x = c, w \\ y = m, i \end{cases} ; \tag{84}$$

$$\partial_{\bar{\varepsilon}_p} \hat{\rho}_{ci} = \partial_{\bar{\varepsilon}_p} \hat{\rho}_{cm}^{tr} + \partial_{\bar{\varepsilon}_p} \hat{\rho}_{ci}^{ac} - \left(\partial_{\bar{\varepsilon}_p} \hat{\rho}_{ci}^{an} + \partial_{\bar{\varepsilon}_p} \hat{\rho}_{ci}^{rm} + \partial_{\bar{\varepsilon}_p} \hat{\rho}_{wi}^{nc}\right) ; \tag{85}$$

$$\partial_{\bar{\varepsilon}_p} \hat{\rho}_{cm} = \partial_{\bar{\varepsilon}_p} \hat{\rho}_{cm}^{gn} + \partial_{\bar{\varepsilon}_p} \hat{\rho}_{ci}^{rm} + \partial_{\bar{\varepsilon}_p} \hat{\rho}_{wi}^{rm}$$
$$- \left(2 \partial_{\bar{\varepsilon}_p} \hat{\rho}_{cm}^{an} + \partial_{\bar{\varepsilon}_p} \hat{\rho}_{ci}^{an} + \partial_{\bar{\varepsilon}_p} \hat{\rho}_{wi}^{an} + \partial_{\bar{\varepsilon}_p} \hat{\rho}_{ci}^{ac} + \partial_{\bar{\varepsilon}_p} \hat{\rho}_{wi}^{ac} + \partial_{\bar{\varepsilon}_p} \hat{\rho}_{cm}^{tr}\right) ; \tag{86}$$

so that,

$$\partial_{\bar{\varepsilon}_p} \hat{\rho}_{cm}^{gn} = M \, c_{cm}^{gn} \frac{\hat{\rho}_{cm}}{\sqrt{\hat{\rho}_{ci} + \hat{\rho}_{wi}}} ; \tag{87}$$

$$\partial_{\bar{\varepsilon}_p} \hat{\rho}_{xy}^{an} = M \, c_{xy}^{an} \, \hat{\rho}_{cm} \, \hat{\rho}_{xy} ; \quad xy = cm, ci, wi ; \tag{88}$$

$$\partial_{\bar{\varepsilon}_p} \hat{\rho}_{xi}^{ac} = M \, c_{xi}^{ac} \sqrt{\hat{\rho}_{xi}} \, \hat{\rho}_{cm} ; \quad x = c, w ; \tag{89}$$

$$\partial_{\bar{\varepsilon}_p} \hat{\rho}_{cm}^{tr} = M \, c_{cm}^{tr} \, \hat{\rho}_{cm}^{3/2} ; \tag{90}$$

$$\partial_{\bar{\varepsilon}_p} \hat{\rho}_{wi}^{nc} = M \, c_{wi}^{nc} \, \hat{\rho}_{ci}^{3/2} \, \hat{\rho}_{cm} ; \tag{91}$$

$$\partial_{\bar{\varepsilon}_p} \hat{\rho}_{xi}^{rm} = M \, c_{xi}^{rm} \, \hat{\rho}_{xi} ; \quad x = c, w ; \tag{92}$$

where $\partial_{\bar{\varepsilon}_p} \equiv \frac{\partial}{\partial \bar{\varepsilon}_p}$ is the partial derivative operator with respect to equivalent plastic strain ($\bar{\varepsilon}_p$); superscripts gn, an, ac, tr, nc and rm respectively denote dislocation generation, annihilation, accumulation, trapping, nucleation and remobilization processes; $c_{xy}^z$ is the constitutive parameter associated with probability amplitude or frequency of occurrence of dislocation process $z$ ($z$ = gn, an, ac, tr, nc, rm) corresponding to dislocations of type $xy$ ($xy$ = cm, ci, wi).

The following equations describe the temperature and strain rate dependencies of constitutive parameters associated with different dislocation processes:

$$\hat{c}_{xy}^z = \left[1 + r_{xy}^z (\hat{T} - 1)^{s_{xy}^z}\right] \dot{\hat{\varepsilon}}_p^{m_{xy}^z} ; \quad \hat{c}_{xy}^z \equiv \frac{c_{xy}^z}{c_{xy0}^z} ; \quad xy = cm, ci, wi ; \tag{93}$$

$$\hat{m}_{xy}^z = \left[1 + r_{zxy}^m (\hat{T} - 1)^{s_{zxy}^m}\right] ; \quad \hat{m}_{xy}^z \equiv \frac{m_{xy}^z}{m_{xy0}^z} ; \quad xy = cm, ci, wi ; \tag{94}$$

where $c_{xy0}^z$ is the reference (at reference temperature and strain rate) material constant associated with probability amplitude of dislocation process $z$ that involves dislocations of type $xy$; $r_{xy}^z$ and $s_{xy}^z$ are respectively temperature sensitivity coefficient and exponent associated with probability amplitude of dislocation process $z$ that involves dislocations of type $xy$; $m_{xy}^z$ and $m_{xy0}^z$ are current and reference (at reference temperature) strain rate sensitivities associated with dislocation process $z$ corresponding to dislocations of type $xy$, respectively; and $r_{zxy}^m$ and $s_{zxy}^m$ are temperature sensitivity coefficient and exponent associated with strain sensitivity of dislocation process $z$ of dislocations of type $xy$, respectively.

Moreover, among dislocation processes, only dislocation generation and accumulation are athermal and rate-independent processes, while the rest of dislocation processes are thermal (temperature-dependent) and (strain) rate-dependent:

$$r_{xy}^z \begin{cases} > 0: & z = \text{an}, \text{tr}, \text{rm}, s\text{pn}, s\text{rm} \\ \gtreqless 0: & z = \text{nc} \\ = 0: & z = \text{gn}, \text{ac} \end{cases} \quad ; \quad s_{xy}^z \begin{cases} > 0: & z = \text{an}, \text{tr}, \text{nc}, \text{rm} \\ = 0: & z = \text{gn}, \text{ac} \end{cases} \quad ; \tag{95}$$

$$m_{xy}^z \begin{cases} < 0: & z = \text{an}, \text{tr} \\ \gtreqless 0: & z = \text{nc}, \text{rm} \\ = 0: & z = \text{gn}, \text{ac} \end{cases} \quad ; \quad xy = cm, ci, wi \,. \tag{96}$$

Given Eqs. (54), (61) and (64), plastic power and generated heat rate due to plastic work are calculated as follows:

$$\dot{w}_p = \bar{\sigma}\,\dot{\bar{\varepsilon}}_p = \sigma_y \dot{\varepsilon}_p \,; \quad \dot{q}_p = \beta \dot{w}_p = \beta \sigma_y \dot{\varepsilon}_p \,; \tag{97}$$

$$\beta = \left( \frac{2\left(\partial_{\bar{\varepsilon}_p}\hat{\rho}_{cm}^{\text{an}} + \partial_{\bar{\varepsilon}_p}\hat{\rho}_{ci}^{\text{an}} + \partial_{\bar{\varepsilon}_p}\hat{\rho}_{wi}^{\text{an}}\right)}{\partial_{\bar{\varepsilon}_p}\hat{\rho}_{cm}^{\text{gn}}} \right)^{\kappa} \,; \quad \kappa > 0 \,; \tag{98}$$

where $\dot{q}_p$ is the volumetric heat generation rate due to plastic work; and $\beta$ is known as dissipation/conversion factor, inelastic heat fraction, efficiency of plastic dissipation, or the Taylor–Quinney coefficient.

The plastic/strain hardening ($\theta$) is obtained by:

$$\theta \equiv \partial_{\bar{\varepsilon}_p}\sigma_y = \partial_{\bar{\varepsilon}_p}\sigma_p = \theta_c + \theta_w \,; \tag{99}$$

$$\theta_x \equiv \partial_{\bar{\varepsilon}_p}\sigma_{px} = \frac{MbG\tilde{\alpha}_x}{2\sqrt{\rho_{xi}}}\partial_{\bar{\varepsilon}_p}\rho_{xi} = \frac{\partial_{\bar{\varepsilon}_p}\hat{\rho}_{xi}}{2\hat{\rho}_{xi}}\sigma_{px} \,; \quad x = c, w \,; \tag{100}$$

where $\theta_x$ is plastic hardening associated with dislocations of type $x$. Further, viscous/strain-rate hardening ($\varphi$) is calculated as follows:

$$\varphi \equiv \partial_{\dot{\bar{\varepsilon}}_p}\sigma_y = \varphi_v + \varphi_p \,; \tag{101}$$

$$\varphi_v \equiv \partial_{\dot{\bar{\varepsilon}}_p}\sigma_v = \frac{m_v}{\dot{\varepsilon}_p}\left[1 + m_v^m \, \breve{\ln}(\dot{\varepsilon}_p)\right]\sigma_v \,; \tag{102}$$

$$\varphi_p \equiv \partial_{\dot{\bar{\varepsilon}}_p}\sigma_p = \varphi_{pc} + \varphi_{pw} \,; \quad \varphi_{px} \equiv \partial_{\dot{\bar{\varepsilon}}_p}\sigma_{px} = \frac{\partial_{\dot{\bar{\varepsilon}}_p}\hat{\rho}_{xi}}{2\hat{\rho}_{xi}}\sigma_{px} = \frac{\partial \bar{\varepsilon}_p}{\partial \dot{\bar{\varepsilon}}_p}\theta_x \,; \quad x = c, w \,; \tag{103}$$

where $\partial_{\dot{\bar{\varepsilon}}_p} \equiv \frac{\partial}{\partial \dot{\bar{\varepsilon}}_p}$ is the partial derivative operator with respect to equivalent plastic strain rate ($\dot{\bar{\varepsilon}}_p$); $\varphi_v$ and $\varphi_p$ are viscous hardening associated with viscous and plastic stresses, respectively; and $\varphi_{px}$ is the viscous hardening associated with plastic stress of type $x = c, w$. Therefore,

$$\varphi = \frac{m_v}{\dot{\varepsilon}_p}\left[1 + m_v^m \, \breve{\ln}(\dot{\varepsilon}_p)\right]\sigma_v + \frac{\partial \bar{\varepsilon}_p}{\partial \dot{\bar{\varepsilon}}_p}\theta \,. \tag{104}$$

The equations related to the constitutive model are numerically integrated in the next section.

## 4. Numerical integration and algorithmic procedure

For finite element implementation, the differential continuum equations presented in sections (2) and (3) must be numerically integrated with respect to time. Thus, the simulation time is discretized to relatively small increments/steps. Consider a (pseudo) time interval $[t^{(n)}, t^{(n+1)}]$, so that $\Delta t^{(n+1)} \equiv t^{(n+1)} - t^{(n)}$ is the time increment at $(n+1)$-th time step. Accordingly,

$$\Delta(\bullet)^{(n+1)} \equiv (\bullet)^{(n+1)} - (\bullet)^{(n)}\ ; \quad (\dot{\bullet})^{(n+1)} \equiv \frac{\Delta(\bullet)^{(n+1)}}{\Delta t^{(n+1)}}\ ; \tag{105}$$

where $(\bullet)$ can be any time-dependent scaler, vector or tensor (of any order) variable; and superscripts $(n)$ and $(n+1)$ respectively represent the value of corresponding time-dependent variable at the beginning and the end of $(n+1)$-th time increment.

Furthermore, it is emphasized that all the tensor variables and equations in this section belong to the corotational/material frame in which the basis system rotates with the material. Hence, calculation of rotation increments, and rotation of corresponding tensors are necessary before the algorithmic procedure provided in this section. Generally, the commercial FE software packages available today, upon user's request, handle the incremental finite rotations and pass the properly rotated stress and strain increment tensors to their user-defined material subroutine. For instance, the incrementally rotated stress and strain increment tensors passed to the user-defined material subroutines of ABAQUS Explicit (VUMAT) and ABAQUS Standard/implicit (UMAT) are based on the Green-Naghdi and Jaumann rates, respectively (ABAQUS, 2014). Moreover, at the end of the time increment computations, FE solver updates the spatial stress tensor ($\boldsymbol{\sigma}^{(n+1)}$) by rotating the corotational stress tensor ($\underline{\boldsymbol{\sigma}}^{(n+1)}$) back to the spatial configuration.

### 4.1. Trial (elastic predictor) step

Firstly, in trial step, it is assumed that the deformation in time increment $[t^{(n)}, t^{(n+1)}]$ is purely elastic:

$$\dot{\lambda}_{\text{trial}}^{(n+1)} = \dot{\bar{\varepsilon}}_{p\ \text{trial}}^{(n+1)} \equiv 0 \quad \Rightarrow \quad \Delta\lambda_{\text{trial}}^{(n+1)} = \Delta\bar{\varepsilon}_{p\ \text{trial}}^{(n+1)} = 0\ ; \tag{106}$$

$$\underline{\dot{\boldsymbol{\varepsilon}}}_{p\ \text{trial}}^{(n+1)} = \mathbf{0} \quad \Rightarrow \quad \Delta\underline{\boldsymbol{\varepsilon}}_{p\ \text{trial}}^{(n+1)} = \mathbf{0}\ ; \tag{107}$$

where subscript trial denotes the trial step. Considering Eq. (69):

$$\underline{\dot{\boldsymbol{\varepsilon}}}_{e\ \text{trial}}^{(n+1)} = \underline{\dot{\boldsymbol{\varepsilon}}}^{(n+1)} \quad \Rightarrow \quad \Delta\underline{\boldsymbol{\varepsilon}}_{e\ \text{trial}}^{(n+1)} = \Delta\underline{\boldsymbol{\varepsilon}}^{(n+1)}\ . \tag{108}$$

Given Eq. (70):

$$\dot{\varepsilon}^{v\ (n+1)} = \dot{\varepsilon}_{\text{trial}}^{v\ (n+1)} = \mathbf{I} : \underline{\dot{\boldsymbol{\varepsilon}}}^{(n+1)} \quad \Rightarrow \quad \Delta\varepsilon^{v\ (n+1)} = \Delta\varepsilon_{\text{trial}}^{v\ (n+1)} = \mathbf{I} : \Delta\underline{\boldsymbol{\varepsilon}}^{(n+1)}\ . \tag{109}$$

Accordingly, given Eqs. (68), (74) and (105), the trial stress tensor is calculated as follows:

$$\underline{\boldsymbol{\sigma}}_{\text{trial}}^{(n+1)} = \underline{\boldsymbol{\sigma}}^{(n)} + \Delta\underline{\boldsymbol{\sigma}}_{\text{trial}}^{(n+1)}\ ; \tag{110}$$

$$\underline{\dot{\boldsymbol{\sigma}}}_{\text{trial}}^{(n+1)} = \mathbb{C}_e : \underline{\dot{\boldsymbol{\varepsilon}}}_{e\ \text{trial}}^{(n+1)} = \mathbb{C}_e : \underline{\dot{\boldsymbol{\varepsilon}}}^{(n+1)} \quad \Rightarrow \quad \Delta\underline{\boldsymbol{\sigma}}_{\text{trial}}^{(n+1)} = \mathbb{C}_e : \Delta\underline{\boldsymbol{\varepsilon}}_{e\ \text{trial}}^{(n+1)} = \mathbb{C}_e : \Delta\underline{\boldsymbol{\varepsilon}}^{(n+1)}\ ; \tag{111}$$

$$\underline{\dot{\boldsymbol{\sigma}}}_{\text{trial}}^{(n+1)} = 2G\underline{\dot{\boldsymbol{\varepsilon}}}^{(n+1)} + \left(K - \frac{2}{3}G\right)\dot{\varepsilon}^{v\ (n+1)}\mathbf{I} \quad \Rightarrow \quad \Delta\underline{\boldsymbol{\sigma}}_{\text{trial}}^{(n+1)} = 2G\Delta\underline{\boldsymbol{\varepsilon}}^{(n+1)} + \left(K - \frac{2}{3}G\right)\Delta\varepsilon^{v\ (n+1)}\mathbf{I}\ . \tag{112}$$

Given Eqs. (73), (75) and (109):

$$\sigma^{h\ (n+1)} = \mathbf{I}:\underline{\boldsymbol{\sigma}}^{(n+1)} = \sigma_{\text{trial}}^{h\ (n+1)} = \mathbf{I}:\underline{\boldsymbol{\sigma}}_{\text{trial}}^{(n+1)}\ ; \tag{113}$$

$$\underline{\boldsymbol{\sigma}}_{\text{trial}}^{d\ (n+1)} = \underline{\boldsymbol{\sigma}}_{\text{trial}}^{(n+1)} - \boldsymbol{\sigma}^{h\ (n+1)}\ ; \quad \boldsymbol{\sigma}^{h\ (n+1)} = \sigma^{h\ (n+1)}\mathbf{I}\ . \tag{114}$$

Furthermore,

$$\underline{\sigma}^{(n+1)} = \underline{\sigma}^{d\ (n+1)} + \sigma^{h\ (n+1)}. \tag{115}$$

Finally, considering Eq. (78), the trial flow direction reads:

$$\mathbf{N}_{\text{trial}}^{(n+1)} = \frac{3}{2} \frac{\underline{\sigma}_{\text{trial}}^{d\ (n+1)}}{\bar{\sigma}_{\text{trial}}^{(n+1)}}. \tag{116}$$

### 4.2. Return mapping (plastic corrector)

Considering Eq. (108), rewriting Eqs. (72) and (73) for the time increment $[t^{(n)}, t^{(n+1)}]$ leads to:

$$\begin{cases} \underline{\dot{\sigma}}^{d\ (n+1)} = 2G\underline{\dot{\varepsilon}}_e^{d\ (n+1)} \\ \underline{\dot{\sigma}}_{\text{trial}}^{d\ (n+1)} = 2G\underline{\dot{\varepsilon}}_{e\ \text{trial}}^{d\ (n+1)} \end{cases} ; \quad \underline{\dot{\varepsilon}}_{e\ \text{trial}}^{d\ (n+1)} = \underline{\dot{\varepsilon}}^{d\ (n+1)} = \underline{\dot{\varepsilon}}_e^{d\ (n+1)} + \underline{\dot{\varepsilon}}_p^{(n+1)}. \tag{117}$$

Consequently,

$$\underline{\dot{\sigma}}_{\text{trial}}^{d\ (n+1)} = \underline{\dot{\sigma}}^{d\ (n+1)} + 2G\underline{\dot{\varepsilon}}_p^{(n+1)} \quad \Rightarrow \quad \Delta\underline{\sigma}_{\text{trial}}^{d\ (n+1)} = \Delta\underline{\sigma}^{d\ (n+1)} + 2G\Delta\underline{\varepsilon}_p^{(n+1)}. \tag{118}$$

Given Eq. (110):

$$\underline{\sigma}_{\text{trial}}^{d\ (n+1)} = \underline{\sigma}^{d\ (n+1)} + 2G\Delta\underline{\varepsilon}_p^{(n+1)}. \tag{119}$$

The following equations represent the incremental forms of Eqs. (76), (77) and (78):

$$\bar{\sigma}^{(n+1)} = \sqrt{\frac{3}{2}} \|\underline{\sigma}^{d\ (n+1)}\| \quad \Rightarrow \quad \bar{\sigma}_{\text{trial}}^{(n+1)} = \sqrt{\frac{3}{2}} \|\underline{\sigma}_{\text{trial}}^{d\ (n+1)}\| ; \tag{120}$$

$$\dot{\bar{\varepsilon}}_p^{(n+1)} = \sqrt{\frac{2}{3}} \|\underline{\dot{\varepsilon}}_p^{(n+1)}\| \quad \Rightarrow \quad \Delta\bar{\varepsilon}_p^{(n+1)} = \sqrt{\frac{2}{3}} \|\Delta\underline{\varepsilon}_p^{(n+1)}\| ; \tag{121}$$

$$\mathbf{N}^{(n+1)} = \frac{3}{2} \frac{\underline{\sigma}^{d\ (n+1)}}{\bar{\sigma}^{(n+1)}} = \frac{\underline{\dot{\varepsilon}}_p^{(n+1)}}{\dot{\bar{\varepsilon}}_p^{(n+1)}} = \frac{\Delta\underline{\varepsilon}_p^{(n+1)}}{\Delta\bar{\varepsilon}_p^{(n+1)}} \quad \Rightarrow \quad \Delta\underline{\varepsilon}_p^{(n+1)} = \frac{3}{2} \frac{\Delta\bar{\varepsilon}_p^{(n+1)}}{\bar{\sigma}^{(n+1)}} \underline{\sigma}^{d\ (n+1)}. \tag{122}$$

Inserting $\Delta\underline{\varepsilon}_p^{(n+1)}$ from Eq. (122) into Eq. (119) results in:

$$\frac{\underline{\sigma}_{\text{trial}}^{d\ (n+1)}}{\bar{\sigma}^{(n+1)} + 3G\Delta\bar{\varepsilon}_p^{(n+1)}} = \frac{\underline{\sigma}^{d\ (n+1)}}{\bar{\sigma}^{(n+1)}}. \tag{123}$$

Given Eq. (120), taking the Euclidian norm of both sides of Eq. (123) leads to:

$$\bar{\sigma}_{\text{trial}}^{(n+1)} = \bar{\sigma}^{(n+1)} + 3G\Delta\bar{\varepsilon}_p^{(n+1)} ; \quad \frac{\underline{\sigma}_{\text{trial}}^{d\ (n+1)}}{\bar{\sigma}_{\text{trial}}^{(n+1)}} = \frac{\underline{\sigma}^{d\ (n+1)}}{\bar{\sigma}^{(n+1)}}. \tag{124}$$

Combining Eqs. (116), (122) and (124) yields:

$$\mathbf{N}^{(n+1)} = \frac{3}{2} \frac{\underline{\sigma}^{d\ (n+1)}}{\bar{\sigma}^{(n+1)}} = \frac{3}{2} \frac{\underline{\sigma}^{d\ (n+1)}}{\sigma_y^{(n+1)}} = \mathbf{N}_{\text{trial}}^{(n+1)} = \frac{3}{2} \frac{\underline{\sigma}_{\text{trial}}^{d\ (n+1)}}{\bar{\sigma}_{\text{trial}}^{(n+1)}}. \tag{125}$$

Since the flow direction and trial flow direction tensors are equal ($\mathbf{N}^{(n+1)} = \mathbf{N}_{\text{trial}}^{(n+1)}$), the yield surface normal is the same for elastic and plastic steps. Therefore, the return mapping in case of associative isotropic $J_2$ plasticity is also known as radial/classical return mapping. Given Eq. (124), the yield function defined by Eq. (54) becomes:

$$\phi^{(n+1)} = \bar{\sigma}^{(n+1)} - \sigma_y^{(n+1)} = \bar{\sigma}_{\text{trial}}^{(n+1)} - \sigma_y^{(n+1)} - 3G\Delta\bar{\varepsilon}_p^{(n+1)}. \tag{126}$$

Thereby, the trial yield function reads:

$$\phi_{\text{trial}}^{(n+1)} = \bar{\sigma}_{\text{trial}}^{(n+1)} - \sigma_{y\,\text{trial}}^{(n+1)}. \tag{127}$$

According to Kuhn-Tucker complementary conditions (Eq. (64)):

$$\phi_{\text{trial}}^{(n+1)} \begin{cases} \leq 0 : & \text{Elastic step} \\ > 0 : & \text{Plastic step} \end{cases}. \tag{128}$$

In return mapping, stress and plastic strain can be updated by linearizing and solving stress and strain residual functions using an iterative method such as Newton-Raphson (NR). Therefore, there are two general types of return mapping:
- stress-based return mapping, in which the nonlinear yield function in case of plastic step is being solved; and
- strain-based return mapping, in which a nonlinear equation for plastic strain increment must be solved.

*4.3. Numerical integration of constitutive model*

Using forward/explicit Euler method for numerical integration of normalized dislocation densities gives:

$$\hat{\rho}_{xy}^{(n+1)} = \hat{\rho}_{xy}^{(n)} + \Delta\hat{\rho}_{xy}^{(n)}; \quad \Delta\hat{\rho}_{xy}^{(n)} = \Delta\bar{\varepsilon}_p^{(n+1)}\, \partial_{\bar{\varepsilon}_p}\hat{\rho}_{xy}^{(n)}; \quad \hat{\rho}_{xy}^{(n=0)} = \hat{\rho}_{xy0}; \quad \begin{cases} x = c, w \\ y = m, i \end{cases}. \tag{129}$$

Likewise, application of backward/implicit Euler method for numerical integration of normalized dislocation densities results in:

$$\hat{\rho}_{xy}^{(n+1)} = \hat{\rho}_{xy}^{(n)} + \Delta\hat{\rho}_{xy}^{(n+1)}; \quad \Delta\hat{\rho}_{xy}^{(n+1)} = \Delta\bar{\varepsilon}_p^{(n+1)}\, \partial_{\bar{\varepsilon}_p}\hat{\rho}_{xy}^{(n+1)}; \quad \hat{\rho}_{xy}^{(n=0)} = \hat{\rho}_{xy0}; \quad \begin{cases} x = c, w \\ y = m, i \end{cases}. \tag{130}$$

In empirical constitutive models where the equivalent accumulated plastic strain is the (mechanical) ISV, it is updated readily by $\bar{\varepsilon}_p^{(n+1)} = \bar{\varepsilon}_p^{(n)} + \Delta\bar{\varepsilon}_p^{(n+1)}$. For fully implicit constitutive integration, backward Euler (Eq. (130)) or other implicit integration methods need to be applied for updating state variables (dislocation densities) which results in a system of coupled nonlinear equations that must be simultaneously solved along with the NR residual function in the return mapping procedure. However, even fully implicit FE simulations of HEVP in complex thermo-mechanical metal forming processes with high geometrical and material nonlinearities often have very low convergence rate. In order to overcome this convergence issue, time increments must be highly reduced. Therefore, in such cases, application of explicit finite element method with semi-implicit integration of constitutive equations is the most efficient approach. Nonetheless, more sophisticated implicit numerical time integration schemes such as generalized midpoint can improve convergence rate, stability, accuracy and performance of the implicit FE analysis (Ortiz and Popov, 1985). Application of implicit numerical integration schemes such as backward Euler and implicit midpoint methods coupled with the consistency approach (Eqs. (65), (66) and (67)) in a fully implicit return mapping scheme will improve the convergence of implicit FE simulations through increasing computation cost of each time increment (de-Borst and Heeres, 2002; Heeres et al., 2002).

Incremental forms of Eqs. (79), (80), (81), (82) and (83) are:

$$\sigma_y^{(n+1)} = \sigma_v^{(n+1)} + \sigma_p^{(n+1)}; \quad \sigma_p^{(n+1)} = \sigma_{pc}^{(n+1)} + \sigma_{pw}^{(n+1)}; \tag{131}$$

$$\sigma_{px}^{(n+1)} = Mb(G\tilde{\alpha}_x)^{(n+1)}\sqrt{\rho_0\, \hat{\rho}_{xi}^{(n+1)}}; \quad x = c, w; \tag{132}$$

$$(G\tilde{\alpha}_x)^{(n+1)} = (G\tilde{\alpha}_x)^{(n)} = G_0\tilde{\alpha}_{x0}\left[1 + r_{\alpha x}^G\left(\hat{T}^{(n)} - 1\right)^{s_{\alpha x}^G}\right]; \quad \hat{T}^{(n)} \equiv \frac{T^{(n)}}{T_0}; \quad x = c, w; \tag{133}$$

$$\sigma_v^{(n+1)} = \sigma_{v0}^{(n+1)} \left(\dot{\hat{\varepsilon}}_p^{(n+1)}\right)^{m_v^{(n+1)}} ; \quad \sigma_{v0}^{(n+1)} = \sigma_{v0}^{(n)} = \sigma_{v00}\left[1 + r_v\left(\hat{T}^{(n)} - 1\right)^{s_v}\right]; \quad \dot{\hat{\varepsilon}}_p^{(n+1)} \equiv \frac{\dot{\varepsilon}_p^{(n+1)}}{\dot{\varepsilon}_0} ; \tag{134}$$

$$m_v^{(n+1)} = m_{v0}\left[1 + r_v^m\left(\hat{T}^{(n)} - 1\right)^{s_v^m}\right]\left(\dot{\hat{\varepsilon}}_p^{(n+1)}\right)^{m_v^m}. \tag{135}$$

Given Eqs. (93) and (94), temperature and strain rate dependencies of material coefficients associated with probability amplitude of various dislocations processes are incrementally calculated according to:

$$c_{xy}^{z\;(n+1)} = c_{xy0}^z\left[1 + r_{xy}^z\left(\hat{T}^{(n)} - 1\right)^{s_{xy}^z}\right]\left(\dot{\hat{\varepsilon}}_p^{(n+1)}\right)^{m_{xy}^{z\;(n+1)}} ; \quad xy = cm, ci, wi ; \tag{136}$$

$$m_{xy}^{z\;(n+1)} = m_{xy}^{z\;(n)} = m_{xy0}^z\left[1 + r_{z_{xy}}^m\left(\hat{T}^{(n)} - 1\right)^{s_{z_{xy}}^m}\right] ; \quad xy = cm, ci, wi . \tag{137}$$

### 4.4. Stress-based return mapping

In stress-based return mapping, the residual function ($R^{(n+1)}$) to be solved ($R^{(n+1)} = 0$) using the NR scheme is usually the same as the yield function:

$$R^{(n+1)} \equiv \phi^{(n+1)} = \bar{\sigma}^{(n+1)} - \sigma_y^{(n+1)} = \bar{\sigma}_{\text{trial}}^{(n+1)} - \sigma_y^{(n+1)} - 3G\Delta\bar{\varepsilon}_p^{(n+1)} . \tag{138}$$

According to Eqs. (127) and (128), in order to check for viscoplastic yielding, $\sigma_{y\;\text{trial}}^{(n+1)}$ must be computed first. In stress-based return mapping, considering Eq. (131):

$$\sigma_{y\;\text{trial}}^{(n+1)} = \sigma_{v\;\text{trial}}^{(n+1)} + \sigma_{p\;\text{trial}}^{(n+1)} ; \quad \sigma_{p\;\text{trial}}^{(n+1)} = \sigma_{pc\;\text{trial}}^{(n+1)} + \sigma_{pw\;\text{trial}}^{(n+1)} . \tag{139}$$

Given Eqs. (106), (129), (130) and (132):

$$\sigma_{px\;\text{trial}}^{(n+1)} = Mb(G\tilde{\alpha}_x)^{(n)}\sqrt{\rho_0\;\hat{\rho}_{xi\;\text{trial}}^{(n+1)}} ; \quad \hat{\rho}_{xi\;\text{trial}}^{(n+1)} = \hat{\rho}_{xi}^{(n)} ; \quad x = c, w . \tag{140}$$

Since the viscous response associated with viscous stress is instantaneous, in order to calculate the trial viscous stress ($\sigma_{v\;\text{trial}}^{(n+1)}$), equivalent plastic strain rate at the beginning of the time increment ($\dot{\hat{\varepsilon}}_p^{(n)}$) is taken into account. However, to avoid a vanishing of the trial viscous stress, for instance, at the beginning of loading (where $\dot{\hat{\varepsilon}}_p^{(n)} = 0$), instead of $\dot{\hat{\varepsilon}}_p^{(n)}$, a minimum equivalent plastic strain rate ($\dot{\varepsilon}_p^{\min}$) determines the trial viscous stress. Accordingly, a corrected equivalent plastic strain rate ($\dot{\hat{\varepsilon}}_{p\;\text{corr}}^{(n)}$) at the beginning of current time increment is adopted:

$$\dot{\hat{\varepsilon}}_{p\;\text{corr}}^{(n)} \equiv \begin{cases} \dot{\hat{\varepsilon}}_p^{(n)} : & \dot{\hat{\varepsilon}}_p^{(n)} > \dot{\varepsilon}_p^{\min} \\ \dot{\varepsilon}_p^{\min} : & \dot{\hat{\varepsilon}}_p^{(n)} \le \dot{\varepsilon}_p^{\min} \end{cases} ; \quad \dot{\varepsilon}_p^{\min} \equiv \xi^{\min}\dot{\varepsilon}_0 ; \quad 0 < \xi^{\min} < 1 . \tag{141}$$

As suggested by Eq. (141), the minimum equivalent plastic strain rate is assumed to be a fraction ($\xi^{\min}$) of the reference strain rate ($\dot{\varepsilon}_0$). In case of having creep or relaxation deformation modes, the fraction $\xi^{\min}$ must be chosen adequately small. Nevertheless, for most of metal forming cases, a value of $10^{-3} \le \xi^{\min} \le 10^{-2}$ is generally recommended. Therefore, given Eqs. (134) and (135):

$$\sigma_{v\;\text{trial}}^{(n+1)} = \sigma_{v00}\left[1 + r_v\left(\hat{T}^{(n)} - 1\right)^{s_v}\right]\left(\frac{\dot{\hat{\varepsilon}}_{p\;\text{corr}}^{(n)}}{\dot{\varepsilon}_0}\right)^{m_{v\;\text{trial}}^{(n+1)}} ; \tag{142}$$

$$m_{v\;\text{trial}}^{(n+1)} = m_{v0}\left[1 + r_v^m\left(\hat{T}^{(n)} - 1\right)^{s_v^m}\right]\left(\frac{\dot{\hat{\varepsilon}}_{p\;\text{corr}}^{(n)}}{\dot{\varepsilon}_0}\right)^{m_v^m} . \tag{143}$$

Thereby,

$$\begin{cases} \sigma_y^{(n+1)} \equiv \breve{\sigma}_y^{(n+1)}\big(T^{(n)}, \mathbf{s}^{(n)}, \Delta\bar{\varepsilon}_p^{(n+1)}, \Delta t^{(n+1)}\big) \\ \sigma_{y\,\text{trial}}^{(n+1)} \equiv \breve{\sigma}_{y\,\text{trial}}^{(n+1)}\big(T^{(n)}, \mathbf{s}^{(n)}, \dot{\bar{\varepsilon}}_{p\,\text{corr}}^{(n)}\big) \end{cases} ; \quad \mathbf{s}^{(n)} \equiv \big\{\rho_{cm}^{(n)}, \rho_{ci}^{(n)}, \rho_{wi}^{(n)}\big\}. \tag{144}$$

As mentioned earlier, in stress-based return mapping, in case of plastic step ($\phi_{\text{trial}}^{(n+1)} > 0$), the nonlinear implicit yield function is taken as the residual function, $R^{(n+1)} \equiv \phi^{(n+1)} = 0$ (Eq. (138)), which must be solved for $\Delta\bar{\varepsilon}_p^{(n+1)}$ using a linearization solving scheme such as iterative Newton-Raphson method. The NR loop starts with an initial guess for $\Delta\bar{\varepsilon}_p^{(n+1)}$. Here, it has been taken from $\dot{\bar{\varepsilon}}_{p\,\text{corr}}^{(n)}$:

$$\Delta\bar{\varepsilon}_{p\,\{k=0\}}^{(n+1)} = \dot{\bar{\varepsilon}}_{p\,\text{corr}}^{(n)}\, \Delta t^{(n+1)} ; \tag{145}$$

where subscript $\{k\}$ is the NR loop index. If the residual function $R_{\{k\}}^{(n+1)}$ is close enough to zero with the specified tolerance $\chi$ (e.g. $\chi = 10^{-6}$), the calculated $\Delta\bar{\varepsilon}_{p\,\{k\}}^{(n+1)}$ is taken as $\Delta\bar{\varepsilon}_p^{(n+1)}$:

$$\Delta\bar{\varepsilon}_p^{(n+1)} = \Delta\bar{\varepsilon}_{p\,\{k\}}^{(n+1)} ; \quad \big|\hat{R}_{\{k\}}^{(n+1)}\big| < \chi ; \quad \hat{R}_{\{k\}}^{(n+1)} \equiv \frac{R_{\{k\}}^{(n+1)}}{\sigma_{y\,\text{trial}}^{(n+1)}} ; \tag{146}$$

where $\hat{R}_{\{k\}}^{(n+1)}$ is the normalized NR residual function at $k$-th NR iteration. Otherwise ($\big|\hat{R}_{\{k\}}^{(n+1)}\big| \geq \chi$), $\Delta\bar{\varepsilon}_{p\,\{k\}}^{(n+1)}$ will be updated iteratively using NR linearization:

$$\Delta\bar{\varepsilon}_{p\,\{k+1\}}^{(n+1)} = \Delta\bar{\varepsilon}_{p\,\{k\}}^{(n+1)} - \left(\frac{\mathrm{d}R_{\{k\}}^{(n+1)}}{\mathrm{d}\Delta\bar{\varepsilon}_{p\,\{k\}}^{(n+1)}}\right)^{-1} R_{\{k\}}^{(n+1)} . \tag{147}$$

Given Eq. (138), Eq. (147) becomes:

$$\Delta\bar{\varepsilon}_{p\,\{k+1\}}^{(n+1)} = \Delta\bar{\varepsilon}_{p\,\{k\}}^{(n+1)} + \frac{R_{\{k\}}^{(n+1)}}{3G + H_{vp\,\{k\}}^{(n+1)}} ; \quad H_{vp\,\{k\}}^{(n+1)} \equiv \frac{\mathrm{d}\sigma_{y\,\{k\}}^{(n+1)}}{\mathrm{d}\Delta\bar{\varepsilon}_{p\,\{k\}}^{(n+1)}} ; \tag{148}$$

where $H_{vp}^{(n+1)}$ is the viscoplastic tangent modulus at the end of the current time increment $(n+1)$. After updating the equivalent plastic strain increment (calculation of $\Delta\bar{\varepsilon}_{p\,\{k+1\}}^{(n+1)}$), again the yield function (NR residual) must be calculated (Eq. (138)); and then the NR loop condition (Eq. (146)) needs to be checked with the updated residual. Given Eqs. (84), (85), (88), (90), (91), (92), (96), (129), (130), (131), (132), (134), (135), (136) and (137):

$$H_{vp}^{(n+1)} \equiv \frac{\mathrm{d}\sigma_y^{(n+1)}}{\mathrm{d}\Delta\bar{\varepsilon}_p^{(n+1)}} = H_v^{(n+1)} + H_p^{(n+1)} ; \tag{149}$$

where $H_v^{(n+1)}$ and $H_p^{(n+1)}$ are viscous and plastic tangent moduli, respectively:

$$H_v^{(n+1)} \equiv \frac{\mathrm{d}\sigma_v^{(n+1)}}{\mathrm{d}\Delta\bar{\varepsilon}_p^{(n+1)}} = \frac{m_v^{(n+1)}\left[1 + m_v^m\,\breve{\ln}(\dot{\bar{\varepsilon}}_p^{(n+1)})\right]}{\Delta\bar{\varepsilon}_p^{(n+1)}}\, \sigma_v^{(n+1)} ; \tag{150}$$

$$H_p^{(n+1)} \equiv \frac{\mathrm{d}\sigma_p^{(n+1)}}{\mathrm{d}\Delta\bar{\varepsilon}_p^{(n+1)}} = \frac{\partial_{\bar{\varepsilon}_p}\hat{\rho}_{ci}^{(n)/(n+1)} + m_{cm}^{\text{tr}\,(n+1)}\partial_{\bar{\varepsilon}_p}\hat{\rho}_{cm}^{\text{tr}\,(n)/(n+1)} - \sum\limits_{\substack{z_{xy}=\text{rm}_{ci}\\ \text{nc}_{wi}}}^{\text{an}_{ci}} m_{xy}^{z\,(n+1)}\partial_{\bar{\varepsilon}_p}\hat{\rho}_{xy}^{z\,(n)/(n+1)}}{2\hat{\rho}_{ci}^{(n+1)}}\, \sigma_{pc}^{(n+1)}$$
$$+ \frac{\partial_{\bar{\varepsilon}_p}\hat{\rho}_{wi}^{(n)/(n+1)} + m_{wi}^{\text{nc}\,(n+1)}\partial_{\bar{\varepsilon}_p}\hat{\rho}_{wi}^{\text{nc}\,(n)/(n+1)} - \sum\limits_{z_{xy}=\text{rm}_{wi}}^{\text{an}_{wi}} m_{xy}^{z\,(n+1)}\partial_{\bar{\varepsilon}_p}\hat{\rho}_{xy}^{z\,(n)/(n+1)}}{2\hat{\rho}_{wi}^{(n+1)}}\, \sigma_{pw}^{(n+1)} . \tag{151}$$

Notice that $\Delta\bar{\varepsilon}_{p\,\{k=0\}}^{(n+1)}$ must not be taken zero (Eq. (150)); otherwise, $H_{vp\,\{k=0\}}^{(n+1)}$ will be undefined. This is the reason behind taking $\Delta\bar{\varepsilon}_{p\,\{k=0\}}^{(n+1)} = \dot{\bar{\varepsilon}}_{p\,\text{corr}}^{(n)}\, \Delta t^{(n+1)} > 0$.

### 4.5. Strain-based return mapping

According to Eqs. (126), (131), (132), (133) and (134), in case of plastic step ($\phi_{\text{trial}}^{(n+1)} > 0$):

$$\dot{\bar{\varepsilon}}_p^{(n+1)} = \dot{\varepsilon}_0 \left(\frac{\sigma_v^{(n+1)}}{\sigma_{v0}^{(n+1)}}\right)^{\frac{1}{m_v^{(n+1)}}} ; \quad \sigma_v^{(n+1)} = \sigma_y^{(n+1)} - \sigma_p^{(n+1)} > 0 ; \quad \sigma_y^{(n+1)} = \bar{\sigma}_{\text{trial}}^{(n+1)} - 3G\Delta\bar{\varepsilon}_p^{(n+1)} . \tag{152}$$

which is an implicit function for $\Delta\bar{\varepsilon}_p^{(n+1)}$. Considering Eq. (105), Eq. (152) can be rearranged as follows to define the residual function in strain-based return mapping:

$$R^{(n+1)} \equiv \Delta\bar{\varepsilon}_p^{(n+1)} - \Delta t^{(n+1)} \dot{\bar{\varepsilon}}_p^{(n+1)} = \Delta\bar{\varepsilon}_p^{(n+1)} - \Delta t^{(n+1)} \dot{\varepsilon}_0 \left(\frac{\sigma_v^{(n+1)}}{\sigma_{v0}^{(n+1)}}\right)^{\frac{1}{m_v^{(n+1)}}} ; \tag{153}$$

which ought to be solved ($R^{(n+1)} = 0$) for $\Delta\bar{\varepsilon}_p^{(n+1)}$ using the iterative NR method.

Furthermore, in strain-based return mapping, considering Eqs. (106), (129), (130), (131) and (132):

$$\sigma_{y\ \text{trial}}^{(n+1)} = \sigma_{p\ \text{trial}}^{(n+1)} = \sigma_{pc\ \text{trial}}^{(n+1)} + \sigma_{pw\ \text{trial}}^{(n+1)} ; \quad \sigma_{px\ \text{trial}}^{(n+1)} = Mb(G\tilde{\alpha}_x)^{(n)}\sqrt{\rho_0\ \hat{\rho}_{xi}^{(n)}} ; \quad x = c, w . \tag{154}$$

In strain-based return mapping, depending on explicit or implicit finite elements, the following initial guess for $\Delta\bar{\varepsilon}_p^{(n+1)}$ is adopted to obtain the highest convergence rate and stability:

$$\Delta\bar{\varepsilon}_{p\ \{k=0\}}^{(n+1)} \equiv \begin{cases} \dot{\bar{\varepsilon}}_p^{(n)}\ \Delta t^{(n+1)} : & \text{Implicit Finite Elements} \\ 0 & : & \text{Explicit Finite Elements} \end{cases} ; \tag{155}$$

If the residual $R_{\{k\}}^{(n+1)}$ is close enough to zero with the specified tolerance $\chi$ (e.g. $\chi = 10^{-6}$), the calculated $\Delta\bar{\varepsilon}_{p\ \{k\}}^{(n+1)}$ is taken as $\Delta\bar{\varepsilon}_p^{(n+1)}$:

$$\Delta\bar{\varepsilon}_p^{(n+1)} = \Delta\bar{\varepsilon}_{p\ \{k\}}^{(n+1)} ; \quad \left|\hat{R}_{\{k\}}^{(n+1)}\right| < \chi ; \quad \hat{R}_{\{k\}}^{(n+1)} \equiv \frac{R_{\{k\}}^{(n+1)}}{\Delta t^{(n+1)}\ \xi^{\text{mean}}\ \dot{\varepsilon}_0} ; \quad \xi^{\text{mean}} > 0 ; \tag{156}$$

where $\xi^{\text{mean}}$ determines the approximate average of equivalent plastic strain rate. Otherwise ($\left|\hat{R}_{\{k\}}^{(n+1)}\right| \geq \chi$), $\Delta\bar{\varepsilon}_{p\ \{k\}}^{(n+1)}$ will be updated iteratively using the NR linearization using Eq. (147), with:

$$\frac{dR_{\{k\}}^{(n+1)}}{d\Delta\bar{\varepsilon}_{p\ \{k\}}^{(n+1)}} = 1 + \frac{3G + H_{p\ \{k\}}^{(n+1)}}{m_v^{(n+1)} \sigma_{v\ \{k\}}^{(n+1)}} \Delta t^{(n+1)} \dot{\bar{\varepsilon}}_p^{(n+1)} = 1 + \frac{3G + H_{p\ \{k\}}^{(n+1)}}{m_v^{(n+1)} \sigma_{v\ \{k\}}^{(n+1)}} \Delta t^{(n+1)} \dot{\varepsilon}_0 \left(\frac{\sigma_{v\ \{k\}}^{(n+1)}}{\sigma_{v0}^{(n+1)}}\right)^{\frac{1}{m_v^{(n+1)}}} . \tag{157}$$

### 4.6. Consistent tangent stiffness operator in implicit finite elements

In implicit finite element method for global linearization, the HEVP consistent/algorithmic tangent stiffness operator/modulus/tensor ($\mathbb{C}^{(n+1)}$) must be computed:

$$\mathbb{C}^{(n+1)} \equiv \frac{\partial \Delta \underline{\sigma}^{(n+1)}}{\partial \Delta \underline{\varepsilon}^{(n+1)}} = 2G_{\text{eff}}^{(n+1)} \mathbb{I} + \left(K - \frac{2}{3}G_{\text{eff}}^{(n+1)}\right) \mathbf{I} \otimes \mathbf{I} + H_{\text{eff}}^{(n+1)} \underline{\mathbf{N}}^{(n+1)} \otimes \underline{\mathbf{N}}^{(n+1)} ; \tag{158}$$

so that,

$$H_{\text{eff}}^{(n+1)} \equiv \frac{4}{3}\left(\frac{G}{1 + \frac{3G}{H_{vp}^{(n+1)}}} - G_{\text{eff}}^{(n+1)}\right) ; \quad G_{\text{eff}}^{(n+1)} \equiv \frac{\bar{\sigma}^{(n+1)}}{\bar{\sigma}_{\text{trial}}^{(n+1)}} G = \frac{\sigma_y^{(n+1)}}{\bar{\sigma}_{\text{trial}}^{(n+1)}} G ; \tag{159}$$

where $H_{\text{eff}}^{(n+1)}$ and $G_{\text{eff}}^{(n+1)}$ are effective/elasto-viscoplastic tangent and shear moduli. In case of elastic step in which $\bar{\sigma}^{(n+1)} = \bar{\sigma}_{\text{trial}}^{(n+1)}$, given Eq. (159), $G_{\text{eff}}^{(n+1)} = G$. Moreover, in elastic domain where the equivalent plastic strain increment tends to zero ($\Delta\bar{\varepsilon}_p^{(n+1)} = 0$), according to Eqs. (149), (150) and (151), the viscoplastic tangent modulus approaches infinity ($H_{vp}^{(n+1)} \to \infty$) which leads to $H_{\text{eff}}^{(n+1)} = 0$. Given Eqs. (35) and (158), this is compatible with the fact that for pure elastic deformation $\underline{\mathbb{C}}^{(n+1)} = \mathbb{C}_e$.

### 4.7. Objective stress update algorithm

The trial step and radial return mapping (elastic predictor-plastic corrector) scheme for objective stress update in associative isotropic $J_2$ plasticity with microstructural constitutive model are summarized in Box 1.

---

**Box 1.** Trial step and radial return mapping (elastic predictor – plastic corrector) scheme for objective stress update in associative isotropic $J_2$ plasticity with microstructural constitutive model.

1) Trial step (elastic predictor):

$$\mathbb{C}_e = 2G\mathbb{I} + \left(K - \frac{2}{3}G\right)\mathbf{I}\otimes\mathbf{I}; \quad K = \frac{2(1+v)}{3(1-2v)}G; \tag{1.1}$$

$$\underline{\boldsymbol{\sigma}}_{\text{trial}}^{(n+1)} = \underline{\boldsymbol{\sigma}}^{(n)} + \mathbb{C}_e : \Delta\underline{\boldsymbol{\varepsilon}}^{(n+1)} = \underline{\boldsymbol{\sigma}}^{(n)} + 2G\Delta\underline{\boldsymbol{\varepsilon}}^{(n+1)} + \left(K - \frac{2}{3}G\right)\Delta\varepsilon^{v\,(n+1)}\mathbf{I}; \quad \Delta\varepsilon^{v\,(n+1)} = \Delta\underline{\boldsymbol{\varepsilon}}^{(n+1)}:\mathbf{I}; \tag{1.2}$$

$$\underline{\boldsymbol{\sigma}}_{\text{trial}}^{d\,(n+1)} = \underline{\boldsymbol{\sigma}}_{\text{trial}}^{(n+1)} - \boldsymbol{\sigma}^{h\,(n+1)}; \quad \boldsymbol{\sigma}^{h\,(n+1)} = \sigma^{h\,(n+1)}\mathbf{I}; \quad \sigma^{h\,(n+1)} = \underline{\boldsymbol{\sigma}}_{\text{trial}}^{(n+1)}:\mathbf{I}; \tag{1.3}$$

$$\phi_{\text{trial}}^{(n+1)} = \bar{\sigma}_{\text{trial}}^{(n+1)} - \sigma_{y\,\text{trial}}^{(n+1)}; \quad \bar{\sigma}_{\text{trial}}^{(n+1)} = \sqrt{\frac{3}{2}}\|\underline{\boldsymbol{\sigma}}_{\text{trial}}^{d\,(n+1)}\|. \tag{1.4}$$

2) Check viscoplastic yielding:

IF $\phi_{\text{trial}}^{(n+1)} \leq 0$, THEN it is an elastic step:

$$\underline{\boldsymbol{\sigma}}^{(n+1)} = \underline{\boldsymbol{\sigma}}_{\text{trial}}^{(n+1)}; \quad \underline{\mathbb{C}}^{(n+1)} = \mathbb{C}_e; \quad \mathbf{s}^{(n+1)} = \mathbf{s}^{(n)}; \quad \Delta\bar{\varepsilon}_p^{(n+1)} = 0. \tag{1.5}$$

3) Return mapping ($\phi_{\text{trial}}^{(n+1)} > 0$): solving the stress or strain-based residual function ($R^{(n+1)} = 0$) for $\Delta\bar{\varepsilon}_p^{(n+1)}$ and $\sigma_y^{(n+1)}$ using the iterative Newton-Raphson method and updating the MSVs.

4) Update the stress and HEVP consistent tangent stiffness operator:

$$\underline{\boldsymbol{\sigma}}^{(n+1)} = \underline{\boldsymbol{\sigma}}^{d\,(n+1)} + \boldsymbol{\sigma}^{h\,(n+1)}; \quad \underline{\boldsymbol{\sigma}}^{d\,(n+1)} = \frac{2}{3}\sigma_y^{(n+1)}\underline{\mathbf{N}}^{(n+1)}; \quad \underline{\mathbf{N}}^{(n+1)} = \frac{3}{2}\frac{\underline{\boldsymbol{\sigma}}_{\text{trial}}^{d\,(n+1)}}{\bar{\sigma}_{\text{trial}}^{(n+1)}} = \frac{\Delta\underline{\boldsymbol{\varepsilon}}_p^{(n+1)}}{\Delta\bar{\varepsilon}_p^{(n+1)}}; \tag{1.6}$$

$$H_{\text{eff}}^{(n+1)} \equiv \frac{4}{3}\left(\frac{G}{1 + \frac{3G}{H_{vp}^{(n+1)}}} - G_{\text{eff}}^{(n+1)}\right); \quad G_{\text{eff}}^{(n+1)} = \frac{\sigma_y^{(n+1)}}{\bar{\sigma}_{\text{trial}}^{(n+1)}}G; \tag{1.7}$$

$$\underline{\mathbb{C}}^{(n+1)} = 2G_{\text{eff}}^{(n+1)}\mathbb{I} + \left(K - \frac{2}{3}G_{\text{eff}}^{(n+1)}\right)\mathbf{I}\otimes\mathbf{I} + H_{\text{eff}}^{(n+1)}\underline{\mathbf{N}}^{(n+1)}\otimes\underline{\mathbf{N}}^{(n+1)}. \tag{1.8}$$

5) Calculation of equivalent plastic strain rate, incremental plastic work and generated heat:

$$\dot{\bar{\varepsilon}}_p^{(n+1)} = \frac{\Delta\bar{\varepsilon}_p^{(n+1)}}{\Delta t^{(n+1)}}; \quad \Delta q_p^{(n+1)} = \beta^{(n)/(n+1)}\Delta w_p^{(n+1)}; \quad \Delta w_p^{(n+1)} = \sigma_y^{(n+1)}\Delta\bar{\varepsilon}_p^{(n+1)}. \tag{1.9}$$

The presented algorithm (Box 1) is programmed as various user-defined material subroutines in ABAQUS Explicit (VUMAT) and ABAQUS Standard/implicit (UMAT) with semi-implicit and fully-implicit constitutive integration schemes using both stress-based and strain-based return mapping algorithms, which are available as supplementary materials to this paper (Motaman, 2019). The overall algorithmic procedure of such implementation is illustrated in the flowchart shown in Fig. 3.

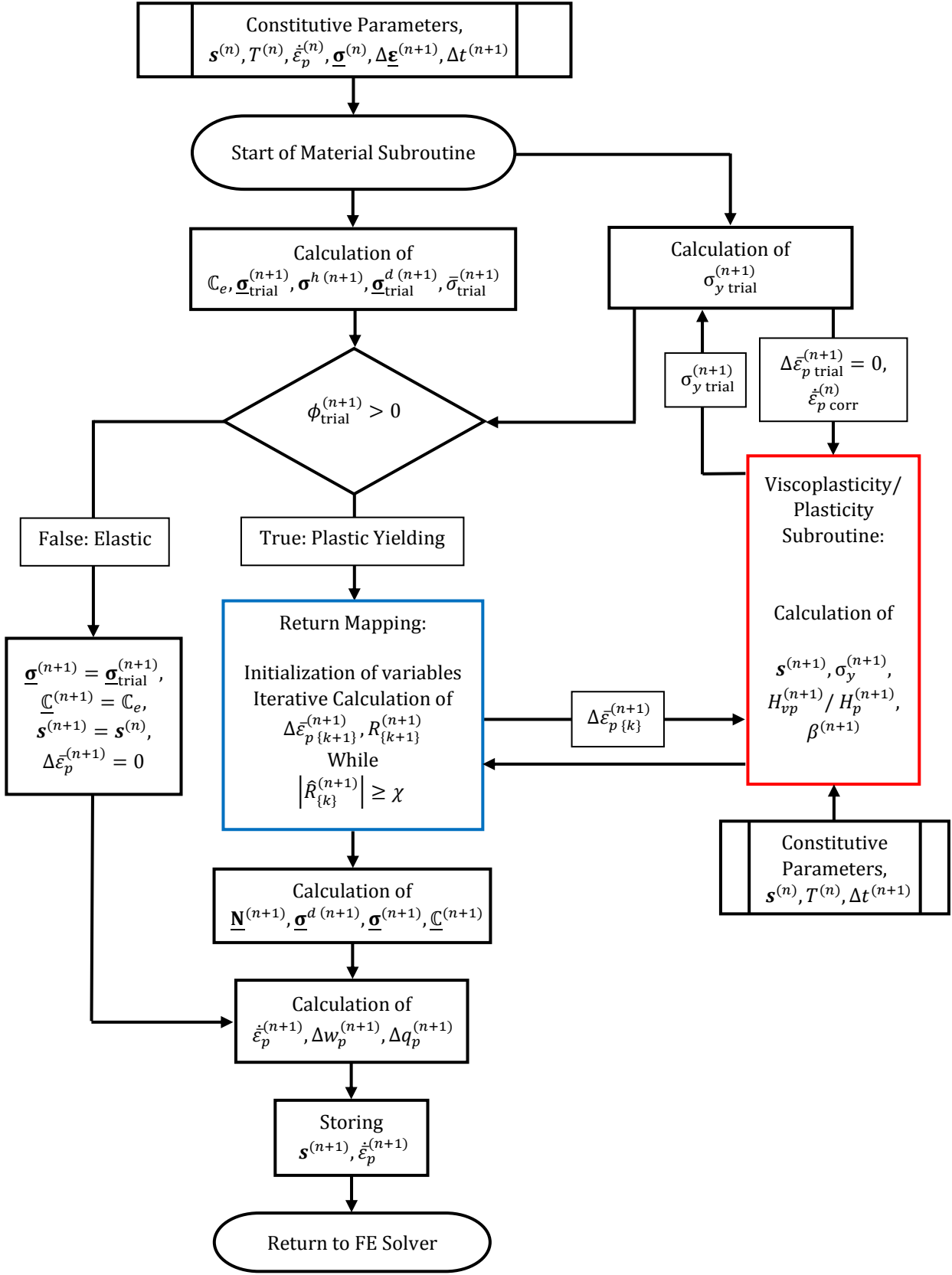

**Fig. 3.** Flowchart illustration of algorithmic procedure for implementation of microstructural material subroutine; the numerically integrated equations of microstructural constitutive model are programmed in the (visco)plasticity subroutine (red box); return mapping loop (blue box), which calls the (visco)plasticity subroutine iteratively, is implemented within the main material subroutine (Motaman, 2019).

# 5. Finite element modeling and simulation

## 5.1. Material and microstructure

The material used in this study is a case-hardenable steel, 20MnCr5 (1.7147, ASI 5120), which is widely used in industrial forging of automotive components such as bevel gears. The chemical composition measured by optical emission spectroscopy (OES) is presented in Table 1.

**Table 1**
Chemical composition of the investigated steel 20MnCr5 [mass%].

| C | Si | Mn | P | S | Cr | Mo | Ni | Cu | Al | N |
|---|---|---|---|---|---|---|---|---|---|---|
| 0.210 | 0.191 | 1.350 | 0.014 | 0.025 | 1.270 | 0.074 | 0.076 | 0.149 | 0.040 | 0.010 |

Furthermore, the microstructure of the undeformed (as-delivered) material consists of equiaxed ferritic-pearlitic grains. Electron backscatter diffraction (EBSD) was used to analyze the microstructure and the texture of undeformed material[*]. The inverse pole figure (IPF) orientation map of the undeformed material sample showing distribution of grain morphology and orientation is demonstrated in Fig. 4.

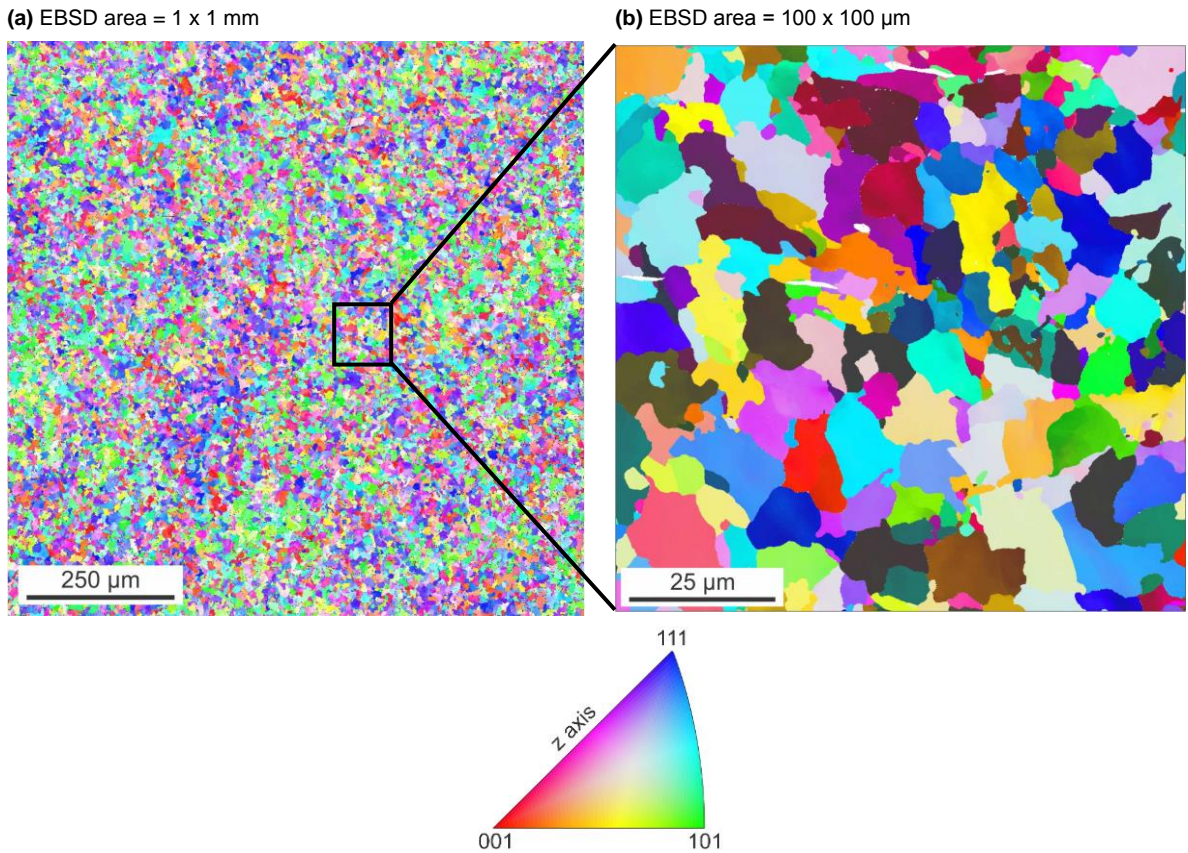

**Fig. 4.** IPF orientation map in the plane (x-y) normal to the symmetry axis (z) of undeformed cylindrical billet. Grain boundaries were identified as boundaries where the misorientation angle is above 5°.

The orientation and grain size distributions are shown in Fig. 5. Pole figures derived from EBSD measurements of a relatively large area in the plane normal to the symmetry axis of undeformed billet for different crystallographic poles/directions are shown in Fig. 5 (a). Furthermore, the grain size distribution calculated based on analysis of EBSD data of the aforementioned large area is shown in Fig. 5 (b). According to Fig. 5 (b), the effective grain size which here is defined as the average of mean grain sizes calculated using distribution of grain size number fraction and area fraction is 8.23 µm, for the investigated material. Furthermore, from the evaluated orientation map (Fig. 4) and pole figures (Fig. 5 (a)), it can be concluded that the undeformed material has a very weak texture (almost random).

---

[*] EBSD measurements were carried out using a a field emission gun scanning electron microscope (FEG-SEM), JOEL JSM 7000F equipped with an EDAX-TSL Hikari EBSD camera. The measurements are conducted at 20 KeV beam energy, approximately 30 nA probe current, and 100 nm step size. OIM software suite (OIM Data Collection and OIM Analysis v7.3) was used to analyze the data.

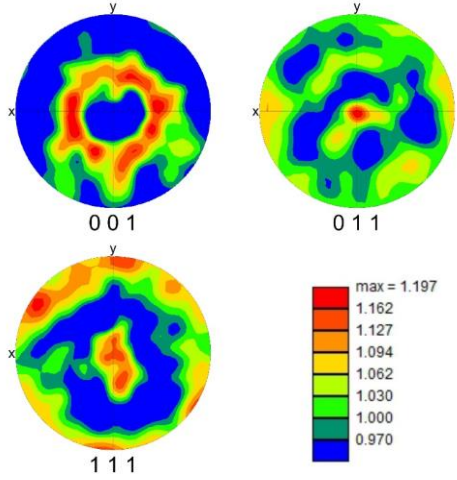
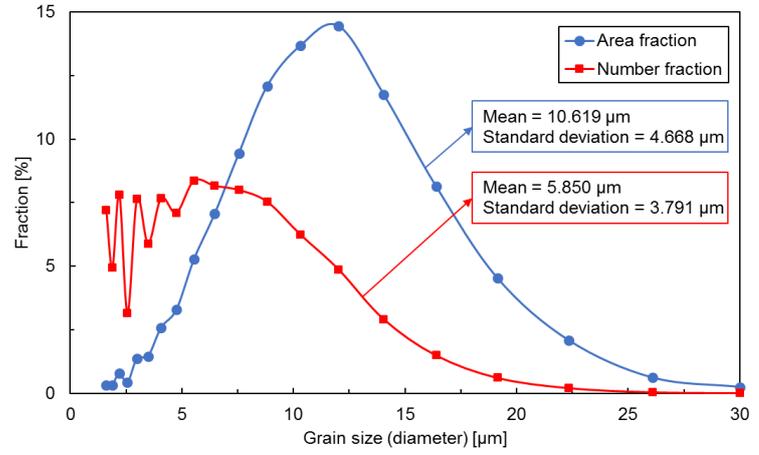

**Fig. 5.** a) pole figures calculated from EBSD measurements (1x1 mm area) in the plane (x-y) normal to the symmetry axis (z) of undeformed cylindrical billet for different crystallographic directions (001, 011 and 111); b) grain size distribution calculated based on analysis of the same EBSD data (the mean and standard deviation values are calculated by fitting to normal/lognormal distribution functions).

The selected reference variables, Taylor factor and Burgers length of the studied material are listed in Table 2.

**Table 2**
Selected reference variables, mean Taylor factor and Burgers length of the investigated material.

| $T_0$ [°C] | $\dot{\varepsilon}_0$ [s⁻¹] | $\rho_0$ [m⁻²] | $M$ [-] | $b$ [m] |
|---|---|---|---|---|
| 20 | 0.01 | $10^{12}$ | 3.0 | $2.55 \times 10^{-10}$ |

In TMM simulation of HEVP, temperature dependent elastic constants (shear modulus and Poisson's ratio) are required as input. The values of elastic constants are calculated using the JMatPro software for the investigated steel with the composition presented in Table 1. The exported temperature-dependent shear modulus and Poisson's ratio versus temperature were fitted using the familiar temperature-dependence relations (Table 3):

$$\hat{G} = 1 + r^G \left(\hat{T} - 1\right)^{s^G} ; \quad \hat{G} \equiv \frac{G}{G_0} ; \quad r^G < 0 ; \quad s^G > 0 ; \tag{160}$$

$$\hat{\upsilon} = 1 + r^\upsilon \left(\hat{T} - 1\right)^{s^\upsilon} ; \quad \hat{\upsilon} \equiv \frac{\upsilon}{\upsilon_0} ; \quad r^\upsilon > 0 ; \quad s^\upsilon > 0 ; \tag{161}$$

where $r^G$ and $s^G$ are temperature sensitivity coefficient and exponent associated with shear modulus ($G$), respectively; $\upsilon$ is the Poisson's ratio; $\upsilon_0$ is the Poisson's ratio at reference temperature; and $r^\upsilon$ and $s^\upsilon$ are temperature sensitivity coefficient and exponent of Poisson's ratio, respectively.

**Table 3**
Elastic constants and their temperature sensitivity.

| $G_0$ [GPa] | $r^G$ [-] | $s^G$ [-] | $\upsilon$ [-] | $r^\upsilon$ [-] | $s^\upsilon$ [-] |
|---|---|---|---|---|---|
| 82.5 | - 0.095 | 1.460 | 0.2888 | 0.0385 | 1.0 |

The micro-mechanical constitutive parameters of the studied material are taken from Motaman and Prahl (2019). Constitutive parameters associated with probability amplitude of different dislocation processes, interaction strengths, initial dislocation densities and reference viscous stress for the investigated material are presented in Table 4. The corresponding temperature sensitivity coefficients and exponents are listed in Table 5. The constitutive parameters associated with strain rate sensitivity of viscous stress together with the parameter controlling the dissipation factor ($\kappa$) are presented in Table 6.

26Thermo-micro-mechanical simulation of bulk metal forming processes

**Table 4**
Reference constitutive parameters associated with probability amplitude of different dislocation processes, reference interaction strengths, initial dislocation densities and reference viscous stress for the investigated material.

| $c_{cm}^{gn}$ [-] | $c_{cm0}^{an}$ [-] | $c_{ci0}^{an}$ [-] | $c_{wi0}^{an}$ [-] | $c_{ci}^{ac}$ [-] | $c_{wi}^{ac}$ [-] | $c_{cm0}^{tr}$ [-] | $c_{wi0}^{nc}$ [-] |
|---|---|---|---|---|---|---|---|
| $6.2970 \times 10^2$ | 0.1492 | 0.0133 | 0.0312 | 0.4989 | 0.1280 | 1.4184 | $1.5534 \times 10^{-3}$ |

| $c_{ci0}^{rm}$ [-] | $c_{wi0}^{rm}$ [-] | $\tilde{\alpha}_{c0}$ [-] | $\tilde{\alpha}_{w0}$ [-] | $\hat{\rho}_{cm0}$ [-] | $\hat{\rho}_{ci0}$ [-] | $\hat{\rho}_{wi0}$ [-] | $\sigma_{v00}$ [MPa] |
|---|---|---|---|---|---|---|---|
| 0.2261 | 0.0217 | 0.1001 | 0.4725 | $2.2573 \times 10^1$ | $2.6427 \times 10^1$ | 0.9234 | 318.84 |

**Table 5**
Temperature sensitivity coefficients and exponents associated with probability amplitude of different dislocation processes, interaction strengths and viscous stress for the studied material.

| $r_{cm}^{an}$ [-] | $r_{ci}^{an}$ [-] | $r_{wi}^{an}$ [-] | $r_{cm}^{tr}$ [-] | $r_{wi}^{nc}$ [-] | $r_{ci}^{rm}$ [-] | $r_{wi}^{rm}$ [-] | $r_{\alpha c}^{G}$ [-] | $r_{\alpha w}^{G}$ [-] | $r_v$ [-] |
|---|---|---|---|---|---|---|---|---|---|
| 0.0547 | 2.0581 | 0.2045 | 3.9680 | 6.1587 | 5.0910 | 2.0631 | - 0.0835 | - 0.0288 | - 0.3376 |

| $s_{cm}^{an}$ [-] | $s_{ci}^{an}$ [-] | $s_{wi}^{an}$ [-] | $s_{cm}^{tr}$ [-] | $s_{wi}^{nc}$ [-] | $s_{ci}^{rm}$ [-] | $s_{wi}^{rm}$ [-] | $s_{\alpha c}^{G}$ [-] | $s_{\alpha w}^{G}$ [-] | $s_v$ [-] |
|---|---|---|---|---|---|---|---|---|---|
| 8.6725 | 0.9988 | 4.0282 | 1.5593 | 4.8075 | 5.5999 | 3.4306 | 2.8735 | 2.5451 | 0.5115 |

**Table 6**
Constitutive parameters associated with strain rate sensitivity of viscous stress and the parameter controlling the dissipation factor.

| $m_{v0}$ [-] | $r_v^m$ [-] | $s_v^m$ [-] | $m_v^m$ [-] | $\kappa$ [-] |
|---|---|---|---|---|
| 0.027 | 0.0785 | 5.0 | 0.0 | 2.0 |

Some thermo-physical material properties of the investigated material including specific heat capacity and thermal conductivity as functions of temperature are calculated using JMatPro software and supplied to the FE model. Moreover, temperature-dependent mass density and thermal expansion coefficient (with respect to room temperature, 20 °C) in cold and warm regimes is measured by dilatometry experiments. Thermo-physical properties of the studied 20MnCr5 steel grade as functions of temperature are plotted in Fig. 6.

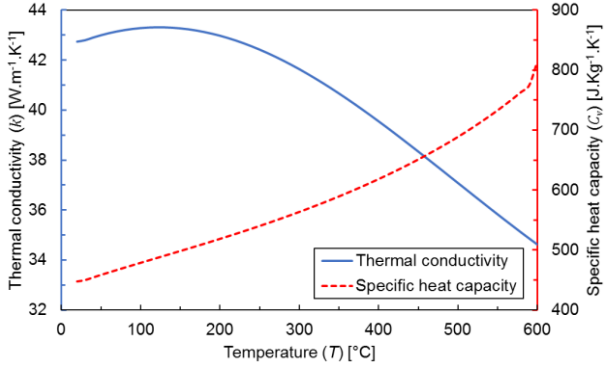
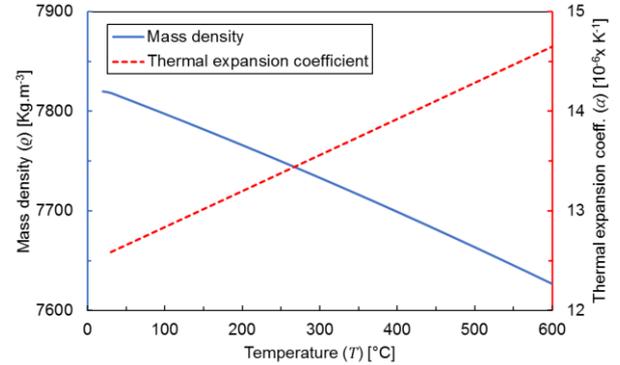
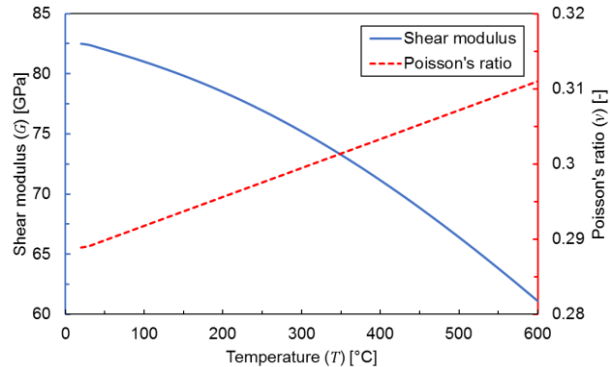

**Fig. 6.** Thermo-physical properties of the of the investigated 20MnCr5 steel grade as functions of temperature.

*5.2. Process*

An industrial warm forging of a bevel gear shaft for automotive applications has been selected as the warm bulk metal forming process to be thermo-micro-mechanically simulated. This process consists of four steps including two forging hits:

1) *preform forging*: the cylindrical forging billet (approximate diameter and length of 54 mm and 112 mm, respectively) is forged in the first forging tool set (punch and die) during 2.5 s. The billet is slightly preheated to about 180 °C (cold regime) just before starting the preform forging operation;
2) *interpass*: this step is the short transfer time (2.5 s) between the end of preform forging and the next forging operation (final forging). The preformed billet which is heated up by preheating as well as adiabatic heating and die-contact friction during preform forging, loses some of the absorbed heat and consequently temperature to the ambient environment mostly due to unforced convection and radiation;
3) *final forging*: after the interpass stage, the somewhat cooled down preformed billet is again forged in the second tool set during 2 s to reach its final shape; and
4) *air cooling*: before performing the subsequent manufacturing processes such as heat treatment and machining on the forged shaft, it is held and consequently reaches the thermal equilibrium at room temperature.

The drawings of radial sections of preformed and (final) forged parts and their images are shown in Fig. 7.

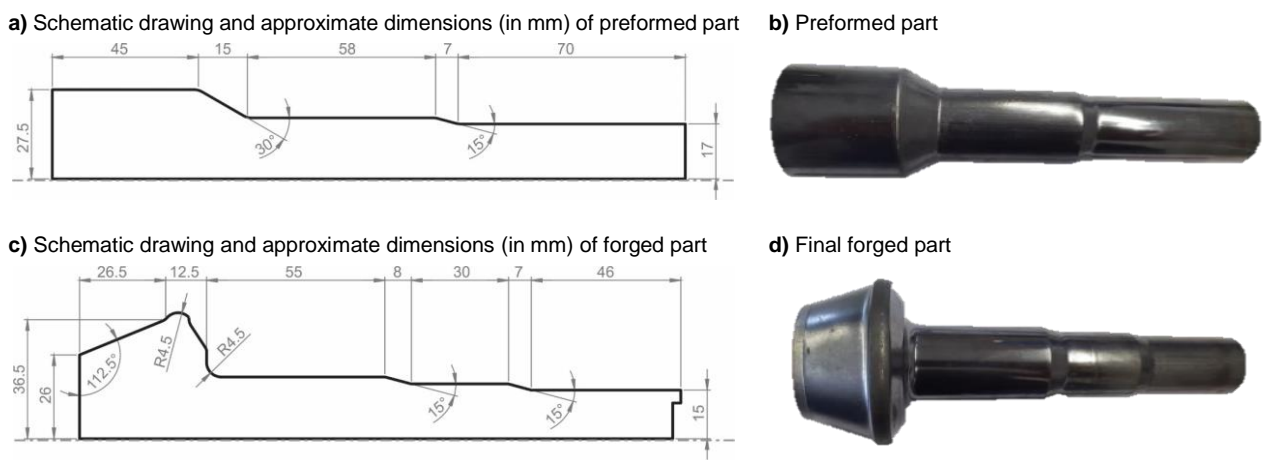

**Fig. 7.** Preformed and forged parts in production of bevel gear.

The following thermo-mechanical boundary conditions are imposed (values of properties are obtained by independent experimental measurements):

- Exploiting axisymmetry of all the parts as well as boundary conditions, only a (two-dimensional) radial section of their assembly is modeled. Thus, appropriate boundary conditions are set to symmetry axes of all parts.
- Similar to its experimental/industrial counterpart, the forging simulation is displacement-controlled. A constant velocity (vertical) of 40 mm.s$^{-1}$ is prescribed to the punches in both deformation steps. There are periods of acceleration and deceleration of punch, respectively, at the beginning and the end of each forging step which last for 0.1 s.
- Constant coulomb-type friction coefficient of 0.05, considering the operation temperature regime and the solid lubricant $MoS_2$ applied on the actual industrial forging (Altan et al., 2004).
- Total generated heat in contact surfaces due to relative motion of contact master and slave surfaces under non-zero (normal) contact pressure is evenly divided between the engaged bodies.
- Thermal contact conductance between the billet and tools as a function of contact pressure and clearance.
- Thermal convection and radiation from forging billet's free surfaces to the ambient environment. Assuming constant convection heat transfer coefficient of 15 Wm$^{-2}$K$^{-1}$ and radiative emissivity coefficient of 0.8 provided an accurate computational prediction of time-temperature loss from a homogenized temperature in warm regime during free cooling in laboratory conditions.

*5.3. Results and validation*

Given the aforementioned material model and properties as well as the introduced process details and assumptions, FE model of the multistep industrial warm forging of bevel gear is created using the ABAQUS CAE software. A fine biased mesh (explicit/implicit 4-node linear thermally coupled axisymmetric, bilinear displacement and temperature) is assigned to the billet. The adaptive Lagrangian-Eulerian remeshing (ALE) algorithm is employed in order to prevent severe element distortion and mesh degradation due to large deformation. Subsequently, the modeled forging process is simulated by the thermo-mechanical/temperature-displacement ABAQUS Explicit solver. Distribution of temperature, MSVs (different types of dislocation density), equivalent stress and equivalent accumulated plastic strain at the end of preform and final forging steps (before unloading) are respectively shown in Fig. 8 and Fig. 9.

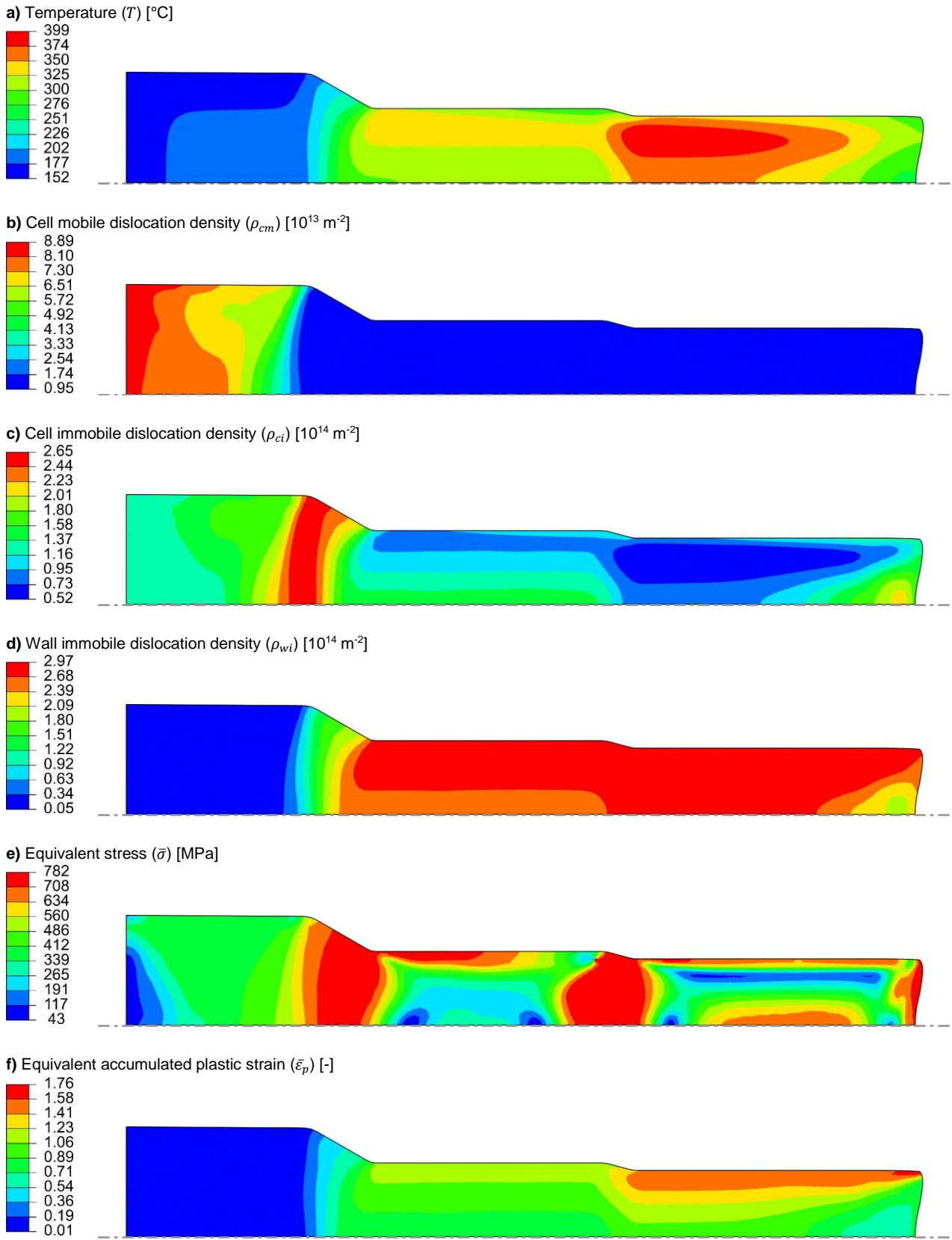

**Fig. 8.** Distribution of temperature, MSVs (different types of dislocation density), equivalent stress and equivalent accumulated plastic strain at the end of preform forging step (before unloading).

S. A. H. Motaman, K. Schacht, C. Haase, U. Prahl    29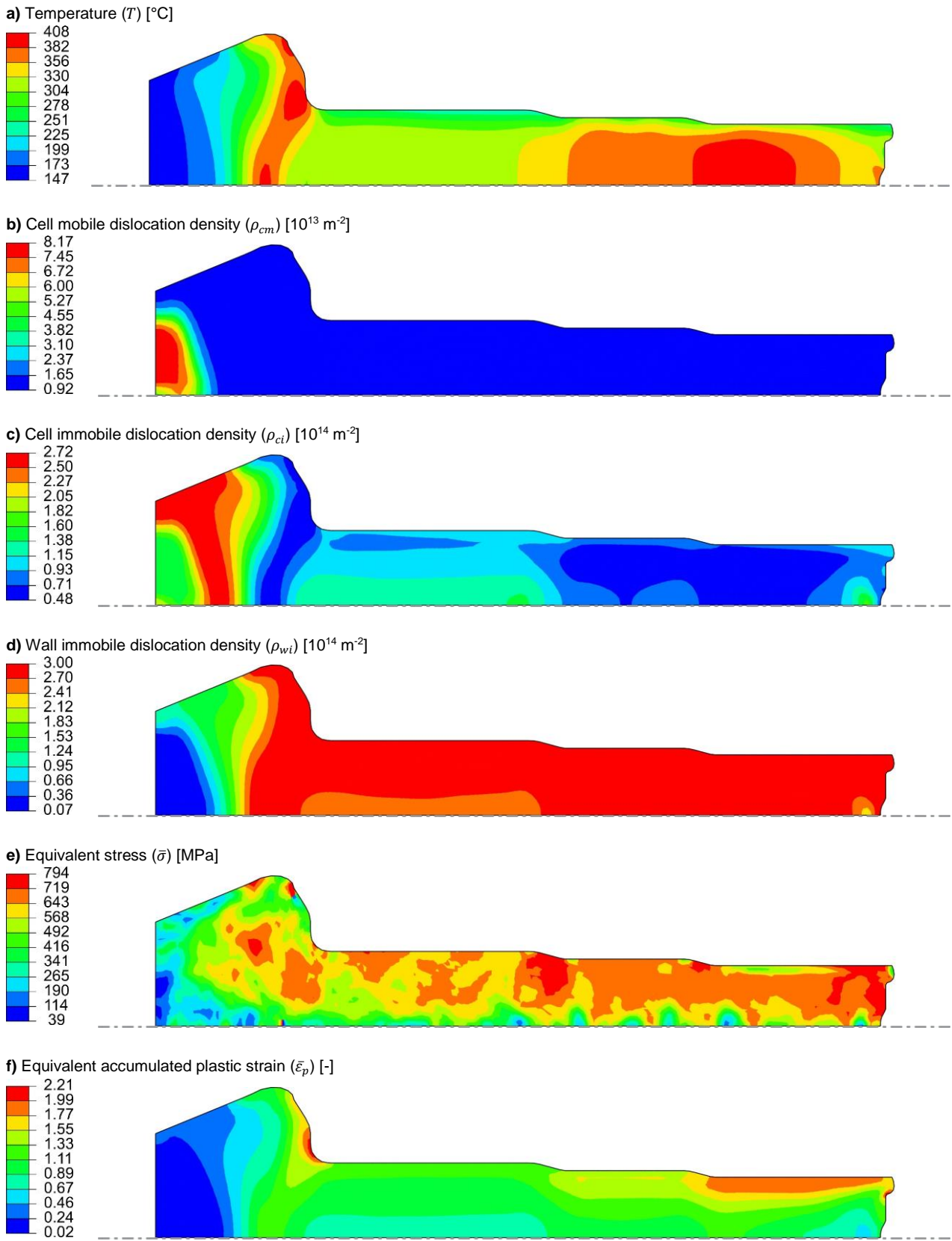

**Fig. 9.** Distribution of temperature, MSVs (different types of dislocation density), equivalent stress and equivalent accumulated plastic strain at the end of final forging step (before unloading).

All the programmed microstructural solvers coupled with their corresponding explicit and implicit thermal and mechanical solvers of ABAQUS have shown a good convergence and stability in TMM-FE simulation of simple uniaxial compression (upsetting) tests. Nonetheless, many trials of explicit and implicit TMM-FE simulations of forging (of bevel gear) with various microstructural solvers revealed that, in case of proper mass scaling, the most efficient and robust microstructural solver is the one with semi-implicit constitutive integration using stress-based return mapping algorithm, implemented as user-defined material subroutines in ABAQUS Explicit (VUMAT). Nevertheless, quantitative comparison of the performance of different integration schemes, and comparison of the results of conventional thermo-mechanical simulations with those of thermo-micro-mechanical simulations are out of scope of the present paper; however, they will make interesting topics for future research.

Although the microstructural constitutive model is validated comprehensively through simple uniaxial compression experiments (Motaman and Prahl, 2019), it still required further validation using experimental deformation under much more complex loading condition (varying temperature, strain rate and stress state) such as the one exists in industrial bulk metal forming processes. TMM-FE simulation of bevel gear is validated by measurement of geometrically necessary dislocation (GND) density ($\rho_{GN}$) using high resolution EBSD, as well as experimental punch force. In order to examine the simulated GND density, which by definition is equal to the wall immobile dislocation density ($\rho_{wi}$), several samples are cut from different regions of the final forged product. From each specimen, an EBSD sample is prepared. Comparison between FE-simulated wall immobile dislocation density and the measured average GND density using EBSD in the final forged part[*], which is manufactured without preheating of the billet prior to preform forging step, is shown in Fig. 10.

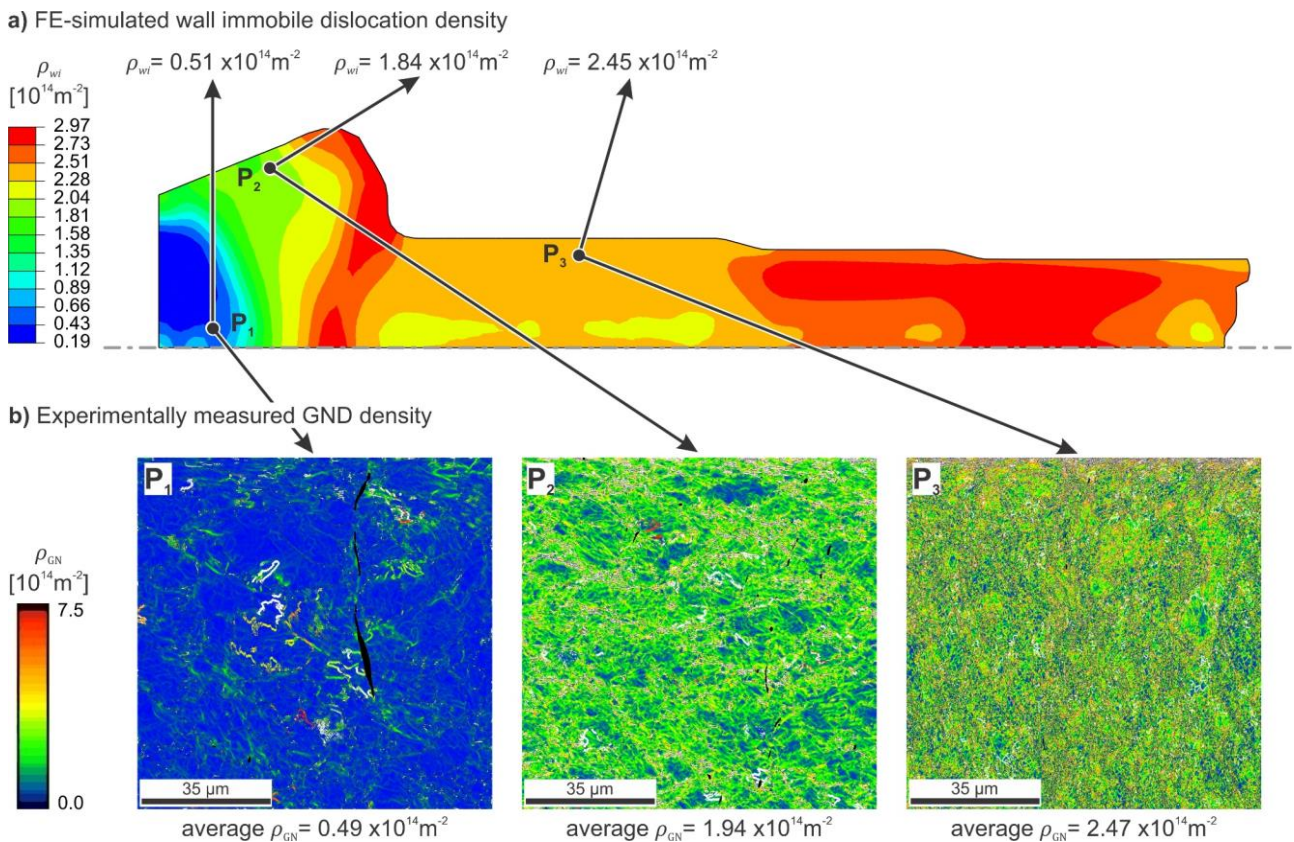

**Fig. 10.** Comparison between FE-simulated wall immobile dislocation density and the experimentally measured average GND density using high resolution EBSD at different points in the final forged part (the billet is not heated prior to preform forging).

Comparison of distribution of wall immobile dislocation density ($\rho_{wi}$) in the final forged parts manufactured with preheating (prior to preform forging) to 180 °C (Fig. 9 (d)) and without preheating (Fig. 10 (a)) reveals that preheating has a significant influence on the distribution of $\rho_{wi}$ and its mean value. In the preheated case, the final forged product has a more homogenous distribution of $\rho_{wi}$; and the mean $\rho_{wi}$ has a larger value. This will result in a more homogenous grain size distribution and finer grains after recrystallization annealing, which is one of the subsequent steps in the manufacturing process chain of the bevel gear. The reason is that $\rho_{wi} = \rho_{GN}$ is the principal driving force for recrystallization because it is the only source of micro-scale residual stresses due to crystal lattice distortions. There is another advantage in the preheating: it lowers the rate of damage accumulation since viscoplastic deformation of ferritic steels in warm regime is followed by a relatively high plastic hardening due to dynamic strain aging (DSA); and generally higher plastic hardening means lower rate of damage accumulation (nucleation and growth of micro-voids).

---

[*] Sample preparation for EBSD involved standard mechanical polishing to 0.05 μm, followed by electropolishing in a 5% perchloric acid and 95% acetic acid solution (by volume) with an applied voltage of 35 V. Measurements are performed using a field emission gun scanning electron microscope (FEG-SEM), JOEL JSM 7000F, at 20 KeV beam energy, approximately 30 nA probe current, and 100-300 nm step size. A Hikari EBSD camera by Ametek-EDAX, in combination with the OIM software suite (OIM Data Collection and OIM Analysis v7.3) by EDAX-TSL, is used for data acquisition and analysis. Subsequently, at each point, GND density is calculated from kernel average misorientation (KAM) which is the average angular deviation between a point and its neighbors in a distance twice the step size as long as their misorientation does not exceed 5°. After mapping KAM values to GND density, over a representative material area with the size of $100 \times 100$ μm, the average GND density is calculated.

The FE-simulated normal force responses of the punches versus time in the preform and final forging steps are compared to their experimental counterparts in Fig. 11.

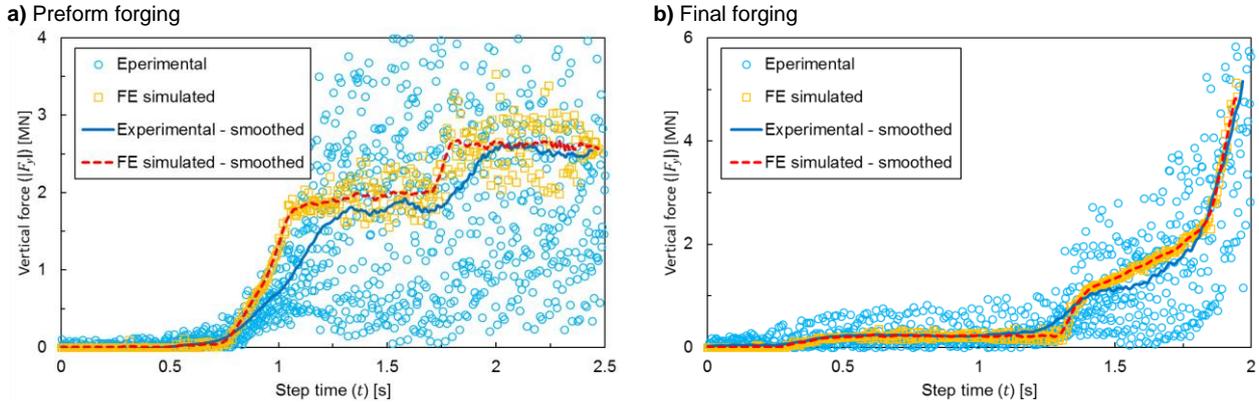

**Fig. 11.** FE-simulated and experimental punch force response versus time in the preform and final forging steps. The unavoidable noise existing in the experimental force plots is associated with the relatively high force and the trade-off between accuracy and stiffness of force measuring devices.

## 6. Concluding remarks

In the introduced method, addition of the microstructural solver, which computes the microstructure/properties evolution, to the main thermal and mechanical solvers enabled fully coupled thermo-micro-mechanical simulation. Since in the cold and warm regimes, (by definition) the microstructure variables are solely the dislocation structures and their associated dislocation densities, by the assumption of isotropy (which is valid for bulk metal forming of initially textureless materials), the state of microstructure of final product and its flow properties as well as the thermo-mechanical aspects of the process were fully determined. The approach proposed and executed in this study has proven to be a sustainable and perhaps the only (computational) solution for comprehensive and simultaneous design of product and process. In summary:

- The theory of continuum finite strain for isotropic hypoelasto-viscoplasticity has been reformulated in the format of rate equations (without using accumulated strain scalars and tensors). This is the only feasible way for correct integration of a microstructural constitutive model based on microstructural state variables (e.g. dislocation densities). Moreover, integration of the microstructural constitutive model using various schemes has been explained in detail.
- The proposed method has shown to be computationally efficient and applicable in industrial scale for optimization of process parameters and tools with respect to properties and microstructure of final products. The cost of TMM implicit FE simulations is higher by orders of magnitude compared to their explicit counterparts. Moreover, the performance of TMM explicit FE simulations with the proposed stress-based return mapping for hypoelasto-viscoplasticity is considerably higher than those performed using strain-based return mapping.
- For the first time, an industrial metal forming process has been thermo-micro-mechanically simulated, and become validated not only by experimental force-displacement but also using measured microstructural state variables, i.e. dislocation density, at different points in the actual final product.

## Acknowledgements

Authors appreciate the support received under the project "IGF-Vorhaben 18531N" in the framework of research program of "Integrierte Umform und Wärmebehandlungssimulation für Massivumformteile (InUWäM)" funded by the German federation of industrial research associations (AiF). The support provided by the project "Laserunterstütztes Kragenziehen hochfester Bleche" from the research association EFB e.V. funded under the number 18277N by AiF is as well gratefully acknowledged. The authors also wish to thank "Schondelmaier GmbH Presswerk" for performing experimental forging.

**Supplementary materials**

Supplementary materials associated with this article are available in the online version as well as the GitHub repository GitHub.com/SAHMotaman/TMM-FE-Simulation:

- Animations of warm forging simulation with demonstration of evolving variables (temperature, equivalent plastic strain rate, stress and MSVs).
- Fortran scripts of the microstructural solver including user-defined material subroutines in ABAQUS Explicit (VUMAT) and ABAQUS Standard/implicit (UMAT) with semi-implicit and fully-implicit constitutive integration schemes using both stress-based and strain-based return mapping algorithms.
- FE models of ABAQUS Standard and ABAQUS Explicit (CAE and INP formats) for uniaxial compression/upsetting test at different (true) strain rates.
- FE models of ABAQUS Standard and ABAQUS Explicit (CAE and INP formats) for cold and warm forging of bevel gear.

S. A. H. Motaman, K. Schacht, C. Haase, U. Prahl    33